\documentclass{aa}  

\usepackage{graphicx}
\usepackage{txfonts}
\usepackage{tikz}
\usepackage{natbib}
\usepackage[colorlinks=true,citecolor=blue]{hyperref}
\usepackage{comment}
\usepackage{pdflscape}
\usepackage{subfig}

\begin{document} 

\title{Disk Evolution Study Through Imaging of Nearby Young Stars (DESTINYS): Characterization of the young star T CrA and its circumstellar environment 
\thanks{Based on observations collected at the European Organisation for Astronomical Research in the Southern Hemisphere under ESO programme 1104.C-0415(H). }}


\author{E.\,Rigliaco\inst{1}
\and R.\,Gratton\inst{1} 
\and S.\, Ceppi\inst{2} 
\and C.\,Ginski.\inst{3,4}  
\and M.\,Hogerheijde\inst{3,4}
\and M.\,Benisty\inst{5,6}
\and T.\,Birnstiel\inst{7,8} 
\and M.\,Dima\inst{1} 
\and S. Facchini\inst{2} 
\and A.\,Garufi\inst{9} 
\and J.\,Bae\inst{10} 
\and M.\,Langlois\inst{11} 
\and G.\,Lodato\inst{2} 
\and E.\,Mamajek\inst{12} 
\and C.F.\,Manara\inst{13} 
\and F.\,M\'enard\inst{14}
\and \'A.\,Ribas\inst{15} 
\and A.\,Zurlo\inst{16,17,18}
}

\institute{INAF/Osservatorio Astronomico di Padova, Vicolo dell'osservatorio 5, 35122 Padova
\email{elisabetta.rigliaco@inaf.it}
\and Dipartimento di Fisica, Universit\`a Degli Studi di Milano, Via Celoria, 16, Milano, 20133, Italy
\and Anton Pannekoek Institute for Astronomy, University of Amsterdam, Science Park 904, 1098XH Amsterdam, The Netherlands 
\and Leiden Observatory, Leiden University, PO Box 9513, 2300 RA, Leiden, The Netherlands
\and Unidad Mixta Internacional Franco-Chilena de Astronom\'{i}a, CNRS/INSU UMI 3386, Departamento de Astronom\'ia, Universidad de Chile, Camino El Observatorio 1515, Las Condes, Santiago, Chile 
\and Univ. Grenoble Alpes, CNRS, IPAG, 38000 Grenoble, France 
\and University Observatory, Faculty of Physics, Ludwig-Maximilians-Universit\"at M\"unchen, Scheinerstr. 1, 81679 Munich, Germany
\and Exzellenzcluster ORIGINS, Boltzmannstr. 2, D-85748 Garching, Germany 
\and INAF, Osservatorio Astrofisico di Arcetri, Largo Enrico Fermi 5, 50125, Firenze, Italy 
\and Department of Astronomy, University of Florida, Gainesville, FL 32611, United States of America 
\and CRAL, UMR 5574, CNRS, Université Lyon 1, 9 avenue Charles André, 69561 Saint-Genis-Laval Cedex, France
\and Jet Propulsion Laboratory, California Institute of Technology, 4800 Oak Grove Drive, Pasadena, CA 91109, USA 
\and European Southern Observatory, Karl-Schwarzschild-Strasse 2, 85748 Garching bei M\"unchen, Germany 
\and Univ. Grenoble Alpes, CNRS, IPAG, 38000 Grenoble, France
\and Institute of Astronomy, University of Cambridge, Madingley Road, Cambridge, CB3 0HA, UK 
\and N\'ucleo de Astronom\'ia, Facultad de Ingenier\'ia y Ciencias, Universidad Diego Portales, Av. Ejercito 441, Santiago, Chile 
\and Escuela de Ingenier\'ia Industrial, Facultad de Ingenier\'ia y Ciencias, Universidad Diego Portales, Av. Ejercito 441, Santiago, Chile 
\and Aix Marseille Univ, CNRS, CNES, LAM, Marseille, France
} 
   \date{Received 12 October 2022; accepted 22 December 2022}


  \abstract
 {In recent years it is emerging a new hot-topic in the star and planet formation field: the interaction between circumstellar disk and its birth cloud. Birth environments of young stars have strong imprints on the star itself and their surroundings. In this context we present a detailed analysis of the wealthy circumstellar environment around the young Herbig Ae/Be star T~CrA.}
 {Our aim is to understand the nature of the stellar system and the extended circumstellar structures as seen in scattered light images. }
 {We conduct our analysis combining archival data, and new adaptive optics high-contrast and high-resolution images.} 
 {The scattered light images reveal the presence of a complex environment around T~CrA composed of a bright forward scattering rim of the disk's surface that is seen at very high inclination, a dark lane of the disk midplane, bipolar outflows, and streamer features likely tracing infalling material from the surrounding birth cloud onto the disk. The analysis of the light curve suggests that the star is a binary with a period of 29.6~years, confirming previous assertions based on spectro-astrometry. 
 The comparison of the scattered light images with ALMA continuum and $^{12}$CO (2–1) line emission shows that the disk is in keplerian rotation, and the northern side of the outflowing material is receding, while the southern side is approaching to the observer. 
 The overall system lays on different geometrical planes.  
 The orbit of the binary star is perpendicular to the outflows and is seen edge on. The disk is itself seen edge-on, with a position angle of $\sim$7$^{\circ}$. 
 The direction of the outflows seen in scattered light is in agreement with the direction of the more distant molecular hydrogen emission-line objects (MHOs) associated to the star.  
 Modeling of the spectral energy distribution (SED) using a radiative transfer scheme well agrees with the proposed configuration, as well as the hydrodynamical simulation performed using a Smoothed Particle Hydrodynamics (SPH) code. 
 }
{
We find evidence of streamers of accreting material around T~CrA. These streamers connect the filament along which T~CrA is forming with the outer parts of the disk, suggesting that the strong misalignment between the inner and outer disk is due to a change in the direction of the angular momentum of the material accreting on the disk during the late phase of star formation. 
This impacts the accretion on the components of the binary, favoring the growth of the primary with respect the secondary, as opposite to the case of aligned disks.
}


\titlerunning{DESTINYS--TCrA}
\authorrunning{Rigliaco et al.}
\maketitle
%

\section{Introduction}

Herbig Ae/Be stars \citep{Herbig1960} are pre-main sequence stars with  intermediate mass covering the range between low-mass T Tauri stars (TTSs) and the embedded massive young stellar objects. The formation of stars in the low and intermediate-mass regimes involves accreting disks formed during the collapse of the protostar, and fast collimated outflows and jets. The circumstellar environment of these objects is highly dynamic and multi-wavelengths observations show large photometric and spectroscopic variability (e.g., \citealt{Pikhartova2021, Mendigutia2011}) that can be used as a tool to understand the physics of accretion and ejection related to the interaction between the star and its circumstellar environment. 

T CrA (RA=19:01:58.79 DEC=-36:57:50.33) is an Herbig Ae/Be star member of the Coronet Cluster, belonging to the Corona Australis star-forming region, which is one of the nearest (149.4$\pm$0.4~pc, \citealt{Galli20}) and most active regions of ongoing star formation. The Coronet Cluster is centered on the Herbig Ae/Be stars R~CrA and T~CrA. 
It is very active in star formation (e.g. \citealt{Lindberg2012}), harboring many Herbig-Haro (HHs) objects and Molecular Hydrogen emission-line Objects (MHOs). It has been target of many surveys, and all studies agree in assigning the Coronet an age $<$3~Myr (e.g. \citealt{Meyer2009, Sicilia-Aguilar2011}). 
In this paper we investigate the variable star T~CrA. 
T~CrA is classified as F0 by \cite{Joy1945}  with effective temperature  T$_{\rm{eff}}$=7200~K, and according to \citet{Cazzoletti2019} and \citet{Herczeg2014} this corresponds to L$_*\sim$29~L$_{\odot}$, and stellar mass $\sim$2.25~M$_{\odot}$ using the evolutionary tracks by \citealt{Siess2000}, and adopting the average distance of 154 pc calculated by \citet{Dzib2018}. 
The Gaia-DR2 and DR3 catalogs \citep{Gaia2016, Gaia2020} do not provide proper motion or parallax for T~CrA. This star was not observed by the Hipparcos satellite and it is also not listed in the UCAC5 catalog. The former UCAC4 catalog \citep{Zacharias2012} provides a proper motion result ($\mu_{\alpha}\cos{\delta}=2.0\pm3.8$~mas~yr$^{-1}$, $\mu_{\delta}$=-22.6$\pm$3.8~mas~yr$^{-1}$), which is consistent with membership in Corona-Australis (within the large uncertainties of that solution). \citet{Galli20} provided an updated census of the stellar population in the Corona Australis deriving an average distance of 149.4$\pm$0.4~pc. This is the distance we will use throughout the paper.  
A deep H$_2$ v=1–0 S(1) 2.12 $\mu$m narrow-band imaging survey of the northern part of the Corona Australis cloud conducted by \citet{Kumar2011} identified many new MHOs \citep{Davis2010}. Among these objects, two are considered unambiguously associated to T~CrA: MHO2013 and MHO2015, see Figure 3 in \citet{Kumar2011}. 
MHO 2015 is a clear bow-shock feature, lying to the south of T~CrA, and it marks the southern lobe of the bipolar outflow originating from T~CrA. MHO 2013 marks the northern lobe. The hypothetical line connecting the two MHOs crosses the position of T~CrA. This is the only unambiguously detected bipolar outflow traced by two complementing bow-shock features in the entire Coronet region \citep{Kumar2011}. We reproduce the image shown in \citet{Kumar2011} in the left panel of Fig.~\ref{fig:sphere_images}. 

T~CrA was suggested to be a binary system by \cite{Bailey1998} and \cite{Takami2003} who adopted spectro-astrometry  in the H$\alpha$ line suggesting that the system is a binary with a companion at $>$0.14$^{\prime\prime}$. However, no companion has been detected using spectro-astrometry in the fundamental rovibrational band of CO at 4.6$\mu$m \citep{Pontoppidan2011} nor with K-band speckle imaging \citep{Ghez1997, Kohler2008}.  
In the same years, infrared speckle observations performed by \cite{Ghez1997} did not show the presence of a stellar companion. The non-detection of the companion by \cite{Ghez1997} implies that the possible companion has a contrast in the K-band larger than 3 mag (that is a K-magnitude fainter than 10.5) or a separation smaller than 0.1 arcsec at the epoch of the observation (April 26, 1994; see also \citealp{Takami2003}). 

Recently, the circumstellar environment of T~CrA has been investigated. SOFIA/FORCAST (Faint Object infraRed CAmera for the SOFIA Telescope, \citealt{Herter2018}) observations show very strong excess in the far-IR. T~CrA was also detected in all Herschel/PACS (Photodetector Array Camera and Spectrometer) bands \citep{Sandell2021}, highlighting the presence of warm or hot dust. 
Mid-infrared interferometric data obtained with VLT/MIDI (MID-infrared Interferometric instrument) show the presence of disk emission from the inner regions, where the temperature is sufficiently high \citep{Varga2018}. 
The presence of the inner disk is also given by the spectral energy distribution (SED) which shows near-IR excess emission \citep{Sicilia-Aguilar2013}. Optical and IR spectra covering the [O{\sc{I}}] $\lambda$6300 and [Ne{\sc II}] 12.81~$\mu$m lines \citep{Pascucci2020} show emission attributed to a jet nearly in the plane of the sky. 
Moreover, continuum ALMA observations of T~CrA at 1.3~mm (230~GHz) were conducted as part of the survey of protoplanetary disks in Corona Australis \citep{Cazzoletti2019} and the data show a $\sim$22$\sigma$ detection at 1.34$^{\prime\prime}$ from the nominal Spitzer position that is considered as detection. The 1.3~mm continuum flux is then converted into a dust mass (M$_{dust}$) under the assumption of optically thin and isothermal sub-millimeter emission, yielding M$_{dust}$=3.64$\pm$0.27~M$_{\oplus}$. No information on the $^{12}$CO(2-1) gas content in the disk are provided. The average disk mass in CrA is 6$\pm$3~M$_{\oplus}$, and it is significantly lower than that of disks in other young (1–3 Myr) star forming regions (Lupus, Taurus, Chamaeleon I, and Ophiuchus) and appears to be consistent with the average disk mass of the 5–10 Myr-old Upper Sco \citep{Cazzoletti2019}. 
 
In this paper we analyze images of T~CrA acquired  with the Very Large Telescope at ESO’s Paranal Observatory in Chile. We employ polarimetric differential imaging (PDI) observations obtained with SPHERE (Spectro-Polarimetric High-contrast Exoplanet REsearch, \citealt{Beuzit2019}) in the H band to explore the circumstellar environment by tracing light scattered by the small ($\mu$m-sized) dust grains. Moreover, we use archival photometric and imaging data to investigate the multiplicity of the system. 
The paper is organized as follows. In Sect.~2 we describe the data collected from the archive and newly acquired. In Sect.~3 we describe the data analysis. First we discuss the multiplicity of the system as suggested by the photometric data, the analysis of the proper motion and the analysis of the PSF subtracted images. Second we analyze the geometry of the system with the analysis of the disk and the extended emission seen in scattered light. In Sect.~4 we propose a scenario that reconciles all the findings, showing a model of the system, and discussing a modeling of the spectral energy distribution and hydrodynamical simulation. In Sect.~5 we summarize and conclude. 


\section{Observations}

\subsection{SPHERE data} 

T~CrA was observed on 2021 June 6th with SPHERE/IRDIS (InfraRed Dual-band Imager and Spectrograph (IRDIS; \citealt{Dohlen2008}) in dual-beam polarimetric imaging mode (DPI; \citealt{deBoer2020, vanHolstein2020}) in the broadband H filter with pupil tracking setting, as part of the DESTINYS program (Disk Evolution Study Through Imaging of Nearby Young Stars, \citet{Ginski2020, Ginski2021}).
An apodized Lyot coronagraph with an inner working angle of 92.5 mas was used to mask the central star. The individual frame exposure time was set to 32~s, and a total of 136 frames were taken separately in 34 polarimetric cycles of the half-wave plate. The total integration time was 72.5 minutes. 
Observing conditions were excellent with an average seeing of 0.8$^{\prime\prime}$ and an atmosphere coherence time of 6.4~ms. 
In addition to the science images, flux calibration images were obtained by offsetting the star position by about 0.5 arcsec with respect to the coronagraph using the SPHERE tip/tilt mirror, and inserting a suitable neutral density filter to avoid image saturation. Two flux calibration sequences were acquired, before and after the science observation. 
We used the public IRDAP pipeline (IRDIS Data reduction for Accurate Polarimetry; \citealt{vanHolstein2020}) to reduce the data. The images were astrometrically calibrated using the pixel scale and true north offset given in \citet{Maire2016}.
Because the data were taken in pupil tracking mode, we were able to perform an angular differential imaging (ADI; \citealt{Marois2006}) reduction in addition to the polarimetric reduction, resulting in a total intensity image and a polarized intensity image. We show the initial combined and flux calibrated Stokes Q and U images as well as the Q$_{\Phi}$ and U$_{\Phi}$ images in Appendix~\ref{sect:AppA}.

\begin{figure*}[h]
\center
\includegraphics[angle=0,width=1.0\textwidth]{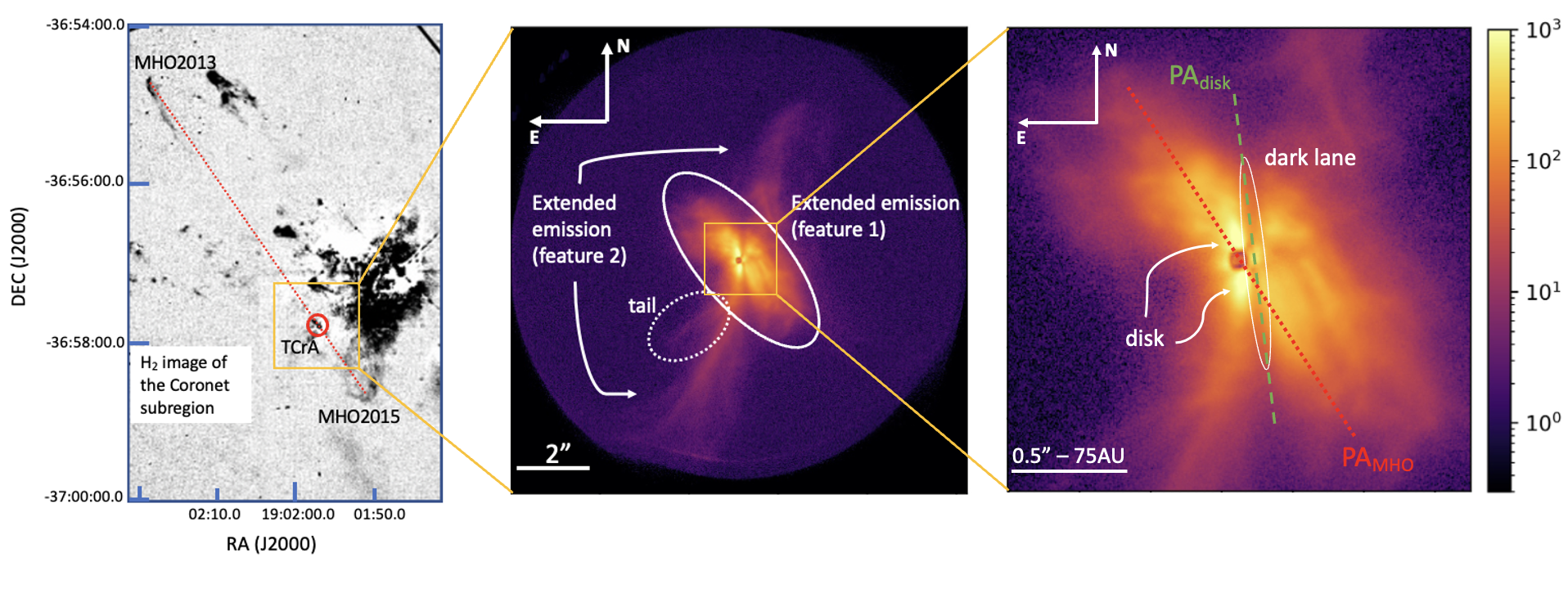}	
\caption{SPHERE/IRDIS polarized light image in H-band of T~CrA. 
{\textit{Left panel:}} H$_2$ image of the Coronet sub-region. The image is adapted from \citealt{Kumar2011}. The red line shows the line connecting the two MHOs associated to T~CrA. The orange box shows the IRDIS field of view. {\it Middle panel}:  Field of view ($\sim$12.5$^{\prime\prime}$) of the SPHERE/IRDIS polarized light image in H-band of T~CrA. The extended emission features analyzed in the manuscript are labeled.  The orange box shows the innermost region of the system. {\textit{Right panel:}}  Zoom-in of the innermost 2$^{\prime\prime}$ around the central system. The disk and the shielded disk mid-plane seen as dark lane are labeled.
} 
\label{fig:sphere_images}
\end{figure*}

Additional SPHERE observations of T~CrA were acquired in 2016 and 2018 with the ESO programs 097.C-0591(A) and 0101.C-0686(A) (P.I. Schmidt) in classical imaging mode, using a classical Lyot coronagraph and the broadband H filter (BB$\_$H). The data were reduced through the SPHERE Data Center \citep{Delorme2017}. The 2016 data have very low S/N ratio and they are not usable for this work. The 2018 IRDIS data are instead of good quality and are used to confirm the features detected in the 2021 images. 

\subsection{NACO data} 

To perform our analysis we also employed archival NACO data. Adaptive optics corrected near-infrared imaging of T~CrA was obtained with NAOS-CONICA (NACO; \citealt{Lenzen2003, Rousset2003}) at the VLT in July 12th 2007 (program ID 079.C-0103(A)), March 29th 2016 (program ID 097.C-0085(A)) and May 21st 2017 (program ID 099.C-0563(A)). 
In all cases images were obtained in Ks band ($\lambda_c$=2.18~$\mu$m) using the S13 camera,  with a 13.72 mas/pixel scale. In 2007, 3000 frames of 0.6 seconds were taken with an average seeing of 0.8. In 2016, 540 frames of 0.5 seconds each were taken with average seeing of 1.5. In 2017, 756 frames of 0.35 seconds each were taken with average seeing of 1.4. The final images are obtained as the median of all the exposures for each year, after re-centering and rotating the single-exposure images. 

\subsection{Photometric data} 
\label{sect:photom_data}

We collected long-term optical photometry of T~CrA from the AAVSO Database\footnote{\url{https://www.aavso.org/data-access}} (American Association of Variable Star Observers: \citealt{Kafka2020}) in order to investigate its secular evolution. 
We also considered data acquired within the ASAS \citep{Pojmanski1997}\footnote{\url{http://www.astrouw.edu.pl/asas/?page=aasc&catsrc=asas3}} and ASAS-SN surveys \citep{Shappee2014}\footnote{\url{https://asas-sn.osu.edu/variables}}. While more accurate than the AAVSO data, they have a much more limited temporal coverage. Results are fully consistent with the long-term light curve obtained from the AAVSO data, but no further insight could be obtained. So we will not discuss the ASAS data further.

\subsection{ALMA data}

T CrA was observed by ALMA on 2016 August 1--2 (project 2015.1.01058.S). Details of the observations and calibration are described in \citet{Cazzoletti2019}. These authors also present an analysis of the continuum emission. For the current paper, the continuum emission was imaged using Hogb\"om CLEANing with Brigss weighting, a robust parameter of 0.5, and a manually drawn CLEAN mask. The resulting beam size is $0.36\times 0.27$ arcsec (PA +78$^\circ$). The noise level is 0.12 mJy, and a continuum flux of 3.1 mJy is detected. These values are not corrected for the primary beam response, which can be expected to affect the results since the observations was not centered on the target. A 2D Gaussian fit to the continuum emission shows that the continuum emission is slightly resolved, with a size of $0.54\times 0.37$ and a PA of +23$^\circ$. 

The $^{12}$CO line emission was imaged using natural weighting and 0.5 km~s$^{-1}$ channels, from $V_{\rm LSR}=-5$ to +15 km~s$^{-1}$; no emission was detected outside this range. We used hand drawn masks for each individual channel and applied multi-scale CLEAN with scales of 0,5,15,25 pixels. A pixel scale of 12.251 mas was used, coincident with the SPHERE pixel scale. Because the CrA region contains extended CO emission around the systemic velocity of T CrA (e.g., \citealt{Cazzoletti2019}), we removed all baselines shorter than 55 k$\lambda$. This removed most, but not all, of the extended line flux but also limits the recovered spatial scales to $\sim 3.75$ arcsec. 

\section{Data Analysis}

The new and archival data described in the previous section allow us to investigate the nature of T~CrA as young stellar object. In this section we will analyze the observational evidences we have for the stellar system, its environment, and the geometry of the extended structures visible in scattered light.  
In Sect.~\ref{sect:binarity} we analyze the clues related to the binarity of the system. In Sect.~\ref{sect:sys_geom} we show the newly acquired polarized light image in H-band of T~CrA, describing all the features that we see in the image. 

\subsection{T~CrA as binary system} 
\label{sect:binarity}

\begin{figure}
\center
\includegraphics[angle=-90,width=0.5\textwidth]{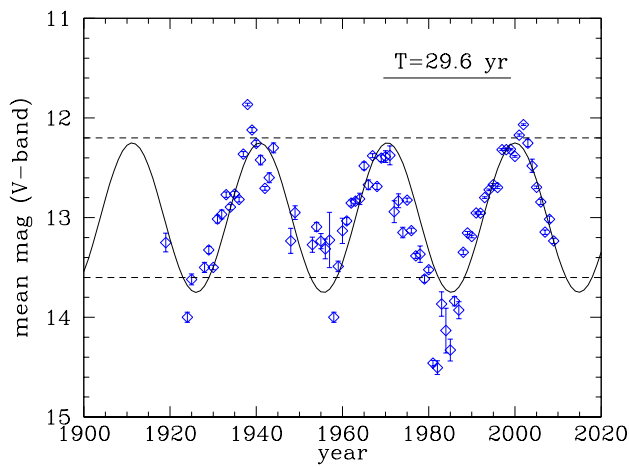} 
\caption{Secular light curve of T~CrA with the photometry collected from the AAVSO archive. Each point is the mean value for each year; error bar is the standard deviation of the mean.The horizontal dashed lines show the $\Delta$V-mag variation. The period of the light curve, measured as the mean between the difference of the first and third maxima and minima, is labeled. 
} 
\label{fig:light_curve}
\end{figure}

The light curve (Fig.~\ref{fig:light_curve}) shows alternate and periodic maxima and minima. 
The photometric time series analyzed in this study consists of more than 5100 V-band data points collected from the AAVSO Database and taken in a period of over 100 years, between 1910 and 2010. Each point in Figure~\ref{fig:light_curve} is the mean value over each year. The secular evolution of the light curve is well reproduced by a sinusoidal function with a period of 29.6 years. 
Sinusoidal light curves, like the one observed in T~CrA, can be due to different reasons such as rotation, pulsation, the presence of eclipsing binaries, or occulting binaries. 
In the case of occulting binaries, the period is generally longer than in the other cases, and the occultation is not only due to the stars, but also to the circumstellar disks surrounding one or both the stars.  
The light curve of T~CrA is suggestive of the motion of an occulting binary star. The variation ($\Delta$V) in V-magnitude is of the order of $\sim$1.4$\pm$0.2 mag (see Fig.~\ref{fig:light_curve}). 

Evidence of the presence of a binary star is also provided by the peculiar proper motion of T~CrA. Indeed T~CrA shows a relative average motion of $7.5\pm 3.8$~mas~yr$^{-1}$ with respect to the cluster in the direction (PA$_{PM}$)=156$\pm$30$^{\circ}$ over the period 1998 (mean epoch of UCAC4 and PPMXL observation) and 2016 (epoch of Gaia DR3).  These values are given by the difference between the proper motion of T~CrA, $\mu_{\alpha}\cos{\delta}=4.2\pm 2.5$~mas~yr$^{-1}$ in RA and $\mu_{\delta}$=-6.2$\pm$2.9~mas~yr$^{-1}$ in DEC (see Appendix~\ref{sect:AppB}), and the average proper motion of the on-cloud Coronet cluster members ($\mu_{\alpha}\cos{\delta}=4.3$~mas~yr$^{-1}$ and $\mu_{\delta}$=-27.3~mas~yr$^{-1}$, \citealt{Galli20}). 
This result might indicate a peculiar (large) motion of T~CrA with respect to the Coronet cluster. However the position of T~CrA is also constrained and defined by the position of the two associated MHOs \citep{Kumar2011}. We measured the position angle of the straight line connecting MHO~2013 and 2015, that are thought to be connected to the star \citep{Kumar2011}, and crossing T~CrA, finding the position angle of the bipolar outflow (PA$_{\rm MHO}$) to be PA$_{\rm MHO}\simeq$33$^{\circ}$. This represents the direction of the large scale bipolar outflows.  
We notice that the minimum distance between T~CrA and the line connecting the two MHOs is only 0.44$^{\prime\prime}$. While this small offset is within the errors in the MHO positions, it can be used to set an upper limit to the relative proper motion of T~CrA with the Coronet cloud in the direction perpendicular to this straight line, that is roughly along the direction where we found an offset between the proper motion of T~CrA measured above and that of the Coronet cluster. The exact value depends on the time elapsed between the expulsion of the material responsible for the MHO and the observation by \citet{Kumar2011}. Given the projected distances from the star of the MHO's are 217$^{\prime\prime}$ (MHO 2013) and 64$^{\prime\prime}$ (MHO 2015), considering the distance of the Coronet cluster and assuming the collimated fast outflowing gas has a speed of approximately 200~km/s as typical for jets from young stars (e.g., \citealt{Frank2014}), we obtain that the material was expelled 765 year ago (for MHO2013) and 224 years ago (for MHO 2015). The upper limit on the proper motion of T~CrA with respect to the cloud is then obtained by dividing the measured offset between the barycenter of the system that includes T Cra and the line connecting the two MHOs: the result is about 1 mas/yr, an order of magnitude less than the offset in proper motions considered above and consistent with the typical scatter of stars in the Coronet cluster. 
We conclude that this offset is not due to a real peculiar motion of T~CrA, that moves as the Coronet cluster, and should then be an apparent or transient effect, that might be due to the orbital motion of the central binary star. 

\begin{figure*}[h]
\center
\includegraphics[angle=0,width=1.0\textwidth]{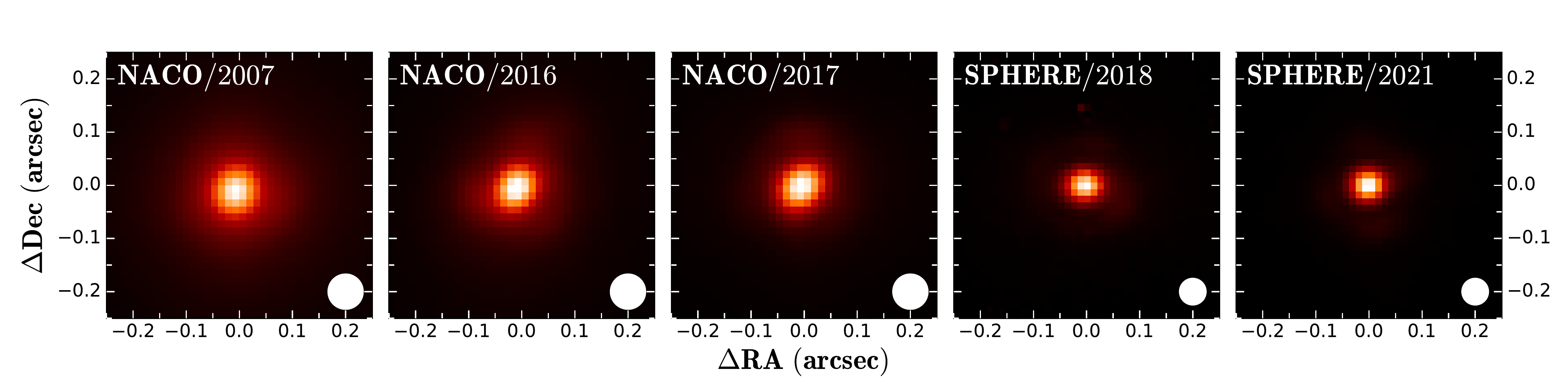} 
\caption{PSF for all the epochs T~CrA was observed. The size of the PSF for every single epochs is shown in the bottom-right corner. For NACO 2016, 2017 data sets we can notice an elongation of the PSF in the NW--SW direction. 
} 
\label{fig:TCrA_binary}
\end{figure*}

Additional evidence of T~CrA as a binary system can also be found in the images acquired with IRDIS in 2018 and 2021 and NACO in 2007, 2016 and 2017. 
We subtracted a median PSF, obtained by rotating and averaging the PSF image in steps of 1 degree, to the raw NACO images taken in 2007 and 2016, 2017. For IRDIS, we used the flux calibration images that are acquired before and after the science sequence. The technique, described by \citet{Bonavita2021}, allows to make a differential image that cancels static aberrations. The output of the procedure is a contrast map that allows to spot stellar companions. 
Due to the contrast limit and to the limits imposed by the diffraction patterns, none of the images obtained allows us to clearly and uniquely detect the presence of a companion star.  However, The PSF of the NACO 2016 and 2017 data set clearly show an extension in the same direction (see Fig.~\ref{fig:TCrA_binary}), namely NW--SE, but in the NACO 2007 data set we do not see this extension. A slight extension can be seen in the SPHERE 2018 data set, while no extension in the SPHERE 2021 data set. The observed extensions, all in the same direction, are very unlikely to be caused by adaptive optic effect, but might indicate a distortion of the PSF due to an unresolved companion.

\subsection{The geometry of the system}  
\label{sect:sys_geom}

Figure~\ref{fig:sphere_images} shows the polarized light image in H-band of T~CrA. 
The image shows several structures, as annotated. In the right panel the brightly illuminated top-side of the outer disk is clearly visible, as well as the shielded disk mid-plane, seen as a stark dark lane in approximately the N-S direction. 
On larger scale, in the middle panel, we can identify two different extended emissions. The extended emission labeled as "feature 1" is two-lobed and extends in the NE--SW direction, up to 2$^{\prime\prime}$ from the central source. The extended emission labeled as "feature 2" appears two lobed as well, it is approximately oriented along the N-S direction. The South lobe extends out to the edge of SPHERE/IRDIS field of view, while the North lobe extends up to $\sim$5$^{\prime\prime}$ from the central source. 
In the following section we will analyze these different structures.

\subsubsection{Outer Disk}
\label{disk_material}

Figure~\ref{fig:sphere_images} in the right panel shows a very prominent morphological feature composed by a dark lane and a bright region that represents the disk surface. This outer disk appears highly inclined, and oriented almost edge-on with respect to the observer, and extends almost to the edge of the coronagraph.  
The dark lane has a maximum width of $\sim$0.2$^{\prime\prime}$ along the E--W direction, corresponding to $\sim$30~au if it were seen exactly edge-on. Moreover, the disk seen as a dark lane shows an offset with respect to the center of the image that corresponds to $\sim$10 pixels in the West direction ($\sim$122~mas) that is about four times the FWHM of the point spread function. 
The disk surface is instead shown by the bright regions that extend further out. 
The PA of the disk measures PA$_{disk}$=7$\pm$2$^{\circ}$, shown as green line in Fig.~\ref{fig:sphere_images}. 
The disk appears highly inclined and seen as a dark lane, as in the case for DoAr25 \citep{Garufi2020}, MY~Lup and IM~Lup \citep{Avenhaus2018}. 
From the images we cannot provide a precise estimate of the disk inclination, but we can make a few considerations. The brightness asymmetry between the bright disk top-side, and the diffuse disk bottom-side, indicates that the disk is not exactly seen edge-on, indeed in that case we should expect top- and bottom-side of the disk to be equally bright. Moreover, the offset between the dark-lane and the center of the image provides another hint of a non-exactly edge-on disk. From simple trigonometric consideration, we can measure the inclination of the disk from the angle between the center of the image and the center of the dark lane and dividing for half the lengths of the dark lane, finding an inclination of $\sim$87$^{\circ}$. We can conservatively assume that the T~CrA disk, identified as a dark lane in the SPHERE image has an inclination between 85-90$^{\circ}$.  
Another possible interpretation for the dark lane could be that it is due to a shadow cast by a highly inclined inner disk close to the center, as in the case of SU~Aur \citep{Ginski2021}. However, in this scenario, we can not reconcile the brightness asymmetry between the bright top-side and the diffuse bottom-side of the disk. Moreover, we should expect the shadow to cross the center of the image, while it appears shifted to the west by $\sim$10~pixels.

\begin{figure}
\center
\includegraphics[angle=0,width=0.5\textwidth]{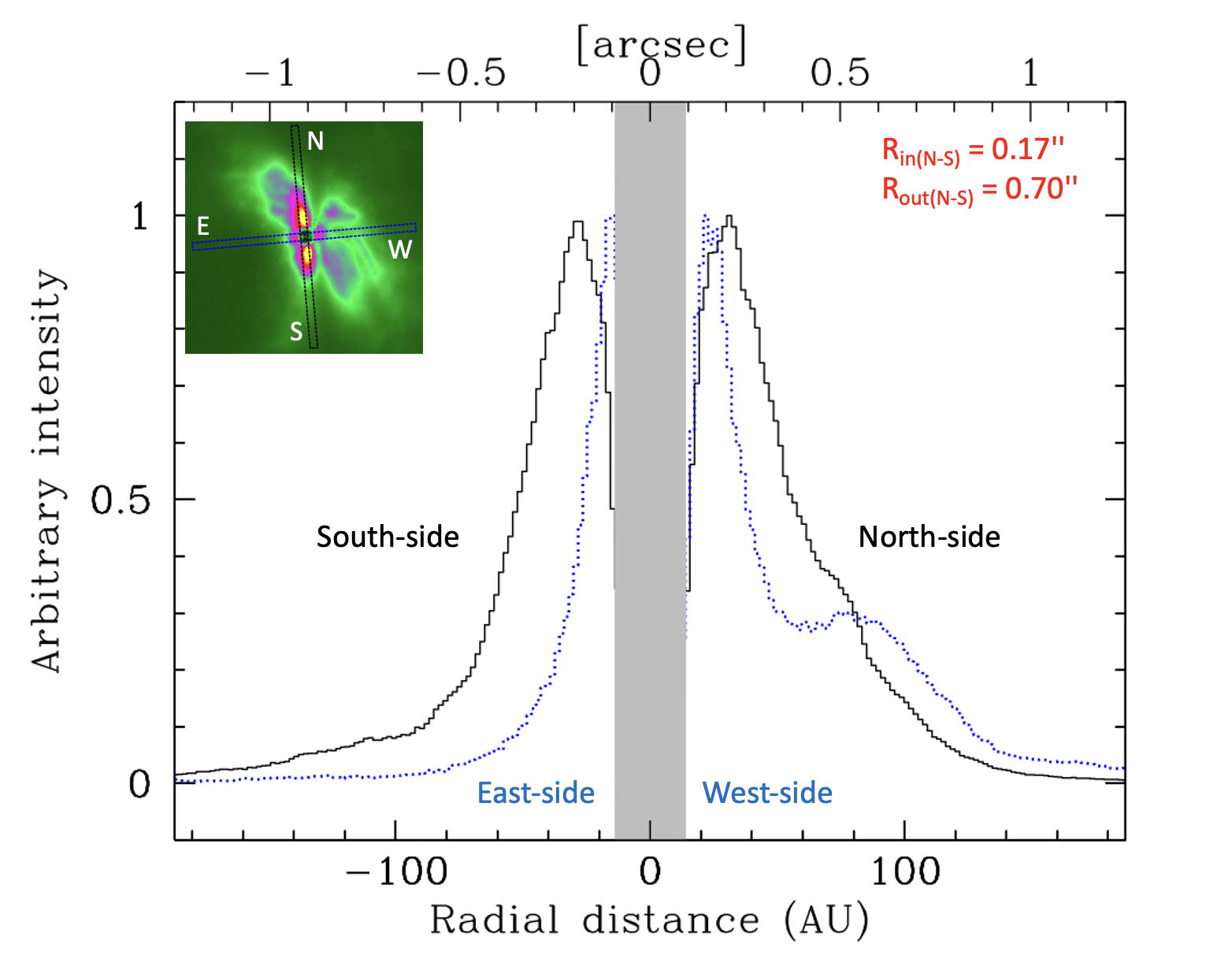}
\caption{Radial profile of the Q$_\phi$ image. The black profile shows the radial profile obtained along a 2.5$^{\prime\prime}$ long slice centered on the star in the N-S direction, with PA=7$^{\circ}$ and extending along the disk (black-dashed box in the insert). The blue-dotted profile shows the radial profile obtained in the orthogonal direction (E-W, blue-dashed box in the insert). All profiles are normalized to the brightness peak of the disk. The gray area shows the radius of the coronagraph. }
\label{fig:combo_rad_prof_TCrA}
\end{figure}

In order to investigate the innermost region of the outer disk, we have plotted the radial profile of the flux seen in Q$_{\Phi}$ scattered light along a slice oriented as the disk, seven pixels wide and 2.5$^{\prime\prime}$ long. The radial profile, normalized to the brightness peak of the disk, is shown in Fig.~\ref{fig:combo_rad_prof_TCrA} as a black line. The grey area shows the coronagraph. The disk has a gap that extends up to $\sim$25~au and is quite symmetric in the innermost region. As far as 60~au the disk start to look asymmetric, and extends up to $\sim$100~au. The observed asymmetry might be due to the outflowing material that overlaps with the disk itself in the north side (as discussed in the next section). From this analysis we consider for the outer disk an inner rim with radius $r_{in}$=0.17$^{\prime\prime}$ ($\sim$25~au) and an outer rim $r_{\rm out}$=0.67$^{\prime\prime}$ ($\sim$100~au). 
We performed the same analysis of the radial profile in the direction orthogonal to the disk, and shown as blue-dotted line in Fig.~\ref{fig:combo_rad_prof_TCrA}. In the East side there is emission from the scattered light down to the border of the coronagraph (r$_{\rm in-east}\lesssim$14~au), and inside the disk rim measured along the disk direction. As expected, in the West-side the emission starts further out, due to the presence of the disk's dark silhouette (r$_{\rm in-west}\sim$30~au). We notice that in the West direction at radial distances $>$50~au there is contamination with the outflowing material. 
We will discuss the presence of scattered light emission inside the outer disk gap in the following section, showing that it may suggest the presence of an intermediate circumbinary disk surrounding the central binary system.  

\subsubsection{Extended emission} 

The structure seen in scattered light in the NE--SW direction, identified as feature 1, is consistent with an outflow in the  direction of the line connecting the two MHOs (MHO2013 and MHO2015) that are unambiguously associated to T~CrA (show in the left panel of Fig.~\ref{fig:sphere_images}), which are however at a projected separation of $\sim$35,000~au and $\sim$10,000~au, respectively. The presence of the MHOs is a clear sign that the source has in the past already experienced outflowing phenomena, hence it is consistent to consider the emission seen in scattered light in the same directions as associated to outflowing material close to the star.
From a geometrical point of view, the dust seen in scattered light in the direction of the outflow has a position angle PA$_{\rm outflow}\sim$35$^{\circ}$ with semi-aperture of $\sim$25$^{\circ}$, consistent with the PA$_{\rm MHO}$ previously discussed. 

The extended emission that elongates approximately in the N-S direction, and identified as feature 2, is two lobed as well. In the North it extends up to 4.5$^{\prime\prime}$ from the center, and appears bent toward the West direction. The Southern feature 2 extends up to the edge of the field of view and appears brighter than the North feature. 
We can also detect a faint dust tail extending from the main disk toward SE. 
As it happens in the case of SU~Aur, where several tails are detected \citep{Ginski2021}, we can trace the tail structure until it merges with the disk. Feature 2 is most likely showing the presence of accretion streamers that bring material from the forming cloud filament to the outer disk.
From the polarized (Fig.~\ref{fig:sphere_images}) and total intensity images of T~CrA 
we can see that in both cases the northern streamer is fainter than the southern streamer, indicating that we overall receive more photons from the South than from the North side of the extended structure. Moreover, the ratio between the polarized and total intensity image shows that the overall degree of polarization is similar on both sides. This indicates that light from the South streamer is scattered with angles smaller than 90$^{\circ}$, favoring the forward scattering. Because the Northern streamer shows a similar degree of polarization, but overall fainter signal, we conclude that the light is scattered with angles larger than 90$^{\circ}$. Hence, the South streamer is angled toward the observed and the North streamer is angled away from the observer.

\section{Discussion}

\begin{figure}[!h]
\center
\includegraphics[angle=0,width=0.45\textwidth]{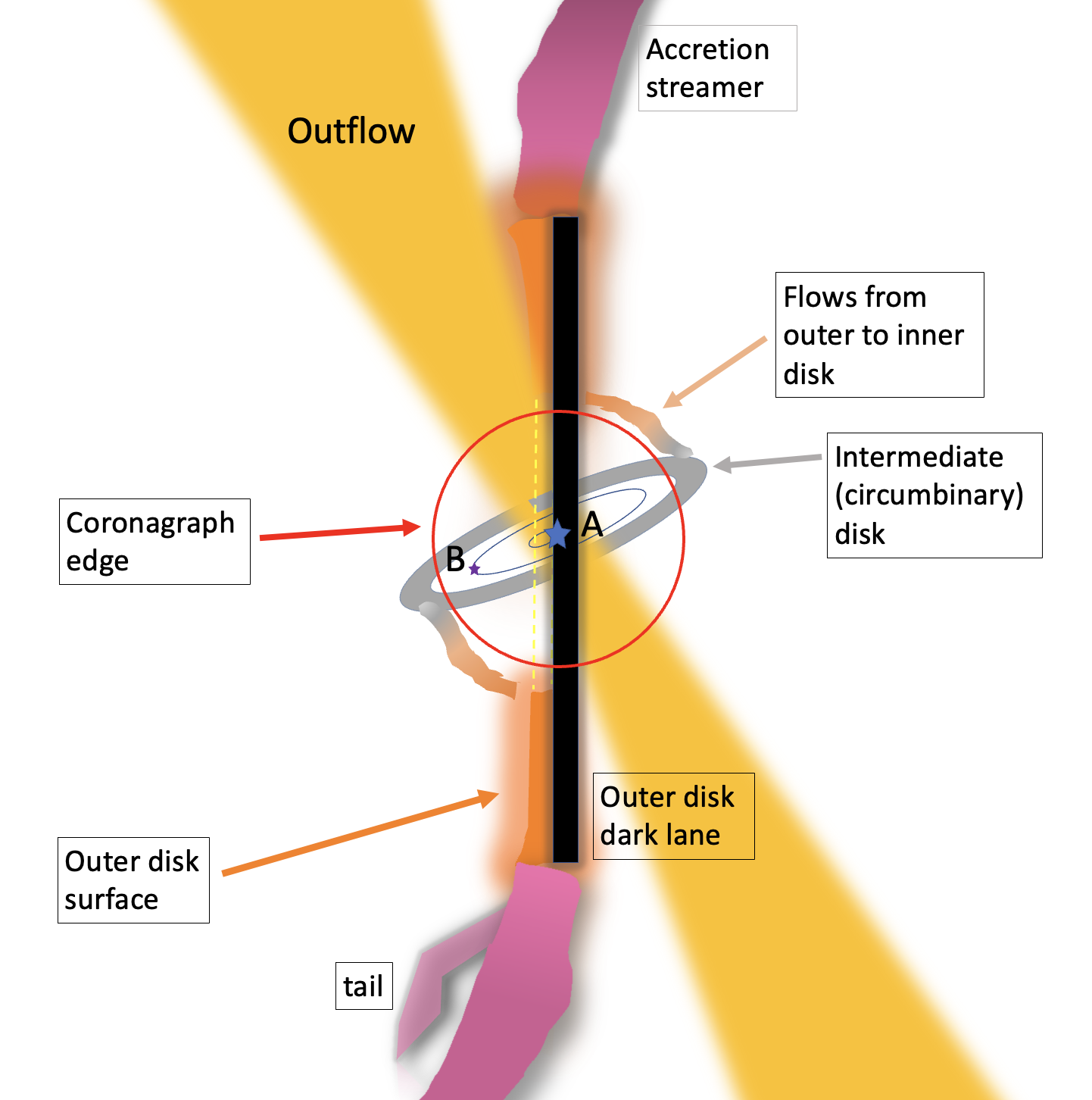}
\caption{Not-to-scale cartoon of the proposed model for the T~CrA system. All the features seen in the scattered light images are labeled. Moreover, the central binary system, and the size of the coronagraph is shown.  }
\label{fig:cartoon}
\end{figure}

The environment around T~CrA is very complex and the analysis of new and archival data shows several features. In the following we will discuss each of the evidences presented in the previous sections, and we will provide a global picture of its circumstellar environment.  
A cartoon of the proposed model, showing all the observational evidences analyzed in the previous section, is shown in Fig.~\ref{fig:cartoon}. 

\subsection{Modeling of the light curve}
\label{sect:mod_lc}

\begin{figure}[!h]
\center
\includegraphics[angle=0,width=0.45\textwidth]{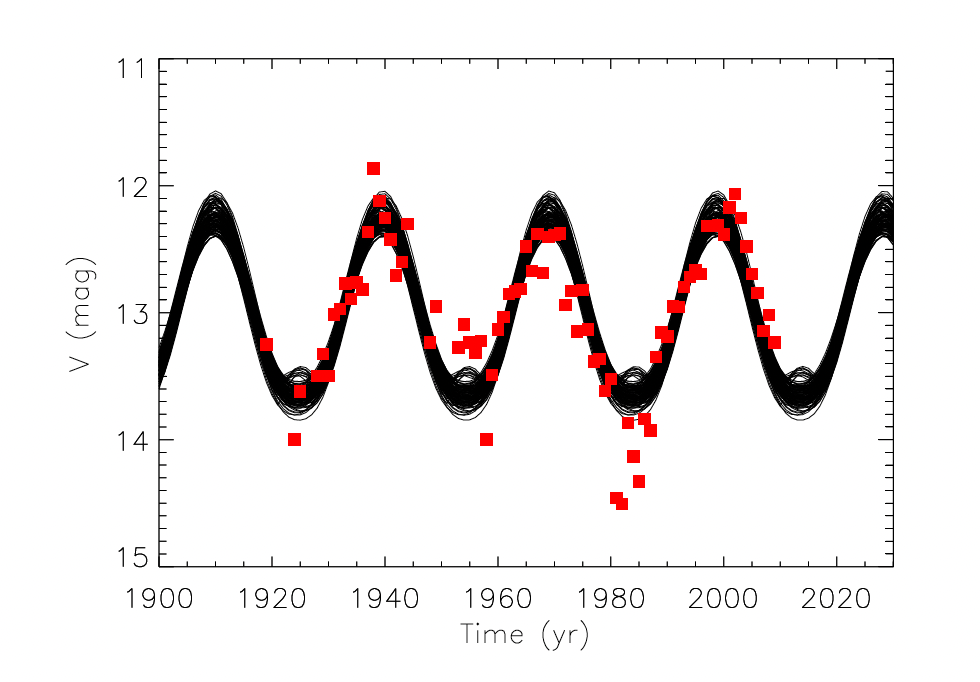} 
\caption{Light curve of T~CrA (red points) compared to the light curves computed with the MC model (black lines) assuming a period of 29.6~years.  }
\label{fig:light_curve_model}
\end{figure}

Motivated by the light curve, the peculiar proper motion and the PSF distortion, we conducted a detailed analysis of the photometric and proper motion data, to be compared to the new information on the system's geometry gathered thanks to the SPHERE's images.  
In the attempt to reproduce the observed light curve and the H-band magnitude collected from 2MASS, we develop a Monte Carlo (MC) model that accounts for the light emitted from a binary system and partially absorbed by a disk seen edge-on, modeled as a slab with an exponential profile, and inclined with respect to the binary's orbit by 35$^{\circ}$, corresponding to an orbit perpendicular to the outflow's direction. For this simplified model we assume for the binary system a circular orbit seen itself edge-on. While the circular orbit is an assumption made to reduce the number of free parameters, and hence avoid degeneracy in the models, the high-inclination of the binary orbit is supported by the observation.  Indeed, as discussed in \citet{Pascucci2020}, evidence from the small blueshift of the [O{\sc{I}}] and [Ne{\sc II}] forbidden lines of T~CrA suggests that the inner disk is itself close to edge-on, with the microjets close to the plane of the sky. 
We assume for the F0 star a mass of 1.7M$_{\odot}$ for the primary star, corresponding to 2~Myrs from the BHAC evolutionary tracks \citep{Baraffe2015}, circular orbit, and a period of 29.6 years as found from the light curve. The model provides the mass ratio ($q$) between the primary and secondary component of the binary system, the epoch of the minimum distance between the two components ($T_0$, in years), the offset of the center of mass with respect to the absorbing slab (disk offset, in mas), the disk thickness (in mas) and the maximum absorption at the disk center (AV$_0$, in mag). The proper motion between the 1998 and 2016 is also measured to be compared to the apparent proper motion of T~CrA.
A corner plot of the derived quantities is shown in Appendix~\ref{sect:AppD}. The MC model computes one million random sampling of the priors, and provides solutions with reduced $\chi^2<$2.3. Figure~\ref{fig:light_curve_model} shows the comparison between the observed secular evolution of T~CrA and the light curve obtained from the model. There is a very good agreement between the observed and modeled light curve. The best fit parameters for each of the computed values, obtained as the median value of all the solutions with $\chi^2<$2.3, are reported in Table~\ref{tab:model_parameters}. 
According to this model T~CrA is a binary system, whose primary star is a 1.7M$_{\odot}$ star, and the secondary is a $\sim$0.9M$_{\odot}$, and it is orbiting with a 29.6~years period. The corresponding semi-major axis of the orbit is $\sim$12~au, seen edge-on, and with the line of nodes of the orbit almost perpendicular to the position angle determined for the outflow. 
Moreover, we check the consistency between the apparent motion as measured from Gaia and ground-based facilities, and the one measured by assuming the motion of the modeled binary system. We find that the offset between the two epochs (1998 and 2016) corresponds to 72$\pm$26~mas, which is consistent with the value of 130$\pm$66~mas measured via Gaia and UCAC4/PPMXL observations, hence justifying the large proper motion of T~CrA with respect to the Coronet motion as due to the motion of the binary system. We will further discuss the results from the model in the next Section.

\begin{table}
\small
\centering
\begin{tabular}{c|c}
\hline
Parameters & Value \\
\hline
log(q) & -0.27$\pm$0.17~M$_{\odot}$ \\ 
T0 & 2006.06$\pm$0.4~years \\
AV$_0$ & 6.7$\pm$1.1~mag \\
Disk Thickness & 54.7$\pm$20.2~mas \\
Disk Offset & 90.7$\pm$19.2~mas \\
\hline
\end{tabular}
\caption{\label{tab:model_parameters} Stellar parameters obtained from the modeling of T~CrA as a binary star. The primary mass star is assumed to be 1.7M$_{\odot}$, the orbit to be circular, and period 29.6 years.}
\end{table}

\subsection{Disk and extended emission} 
\label{sect:disk_structure}

Thanks to the new images acquired with SPHERE/IRDIS, and to the wealth of literature data on this target, we have now a better knowledge of the disk and extended structure around T~CrA, and it appears very composite.  
The disk itself is composed by inner (circumstellar) disk(s) surrounding the primary (secondary) star of the binary system, an intermediate (circumbinary) disk, slightly visible in scattered light, and an outer (circumbinary) disk that is the most prominent in scattered light. 
Together with the extended emission features, we will discuss all these features in the following subsections. 

{\bf Disks.} 
The {\it outer} disk around T~CrA is not continuous. The scattered light images and the radial profile analysis of the Q$_{\Phi}$ image show that the bright top-side of the outer disk extends up to $\sim$100~au in the N-S direction, and show a gap in the same direction that extends down to $\sim$25~au. 

Evidence of an inner (circumstellar) disk(s) surrounding the primary (secondary) star of the binary system comes from the several tracers of gas and dust well beyond the dust gap. \citet{Pascucci2020} analyze the [O{\sc{I}}] $\lambda$6300 and [Ne{\sc II}] 12.81~$\mu$m emission lines observed in high-resolution optical and infrared spectra, and conclude that they are associated to fast and collimated microjets. In addition, the presence of gas can also be inferred from the non-negligible level of mass accretion rate ($\dot{M}_{acc}\sim$8.1$\times$10$^{-9}$~M$_{\odot}$/yr, \citealt{Dong2018, Takami2003}). This gas is most likely distributed into an {\it inner} circumstellar disk, that allows accretion onto the system. 
The presence of the inner disk is also highlighted by mid-infrared interferometric data of the thermal emission of disk \citep{Varga2018}, and by the SED \citep{Sicilia-Aguilar2013,Sandell2021}. 

The images acquired with SPHERE show the presence of scattered light down to the edge of the coronagraph in the E-W direction. The origin of such emission, highly inclined with respect to the outer disk, is not clear. However, as we will see in section~\ref{sect:SPM_model},  it might be due to an {\it intermediate} circumbinary disk, that is a natural transient consequence of the breaking of the innermost circumstellar disks due to the different inclination of inner and outer disks. Evidence of emission very close to the coronagraph edges are also found by \citealt{Cugno2022} using the NaCo imager with the L$^{\prime}$ filter ($\lambda$=3.6~$\mu$m) within the NaCo-ISPY large program.

\begin{figure}
\includegraphics[angle=0,width=0.5\textwidth]{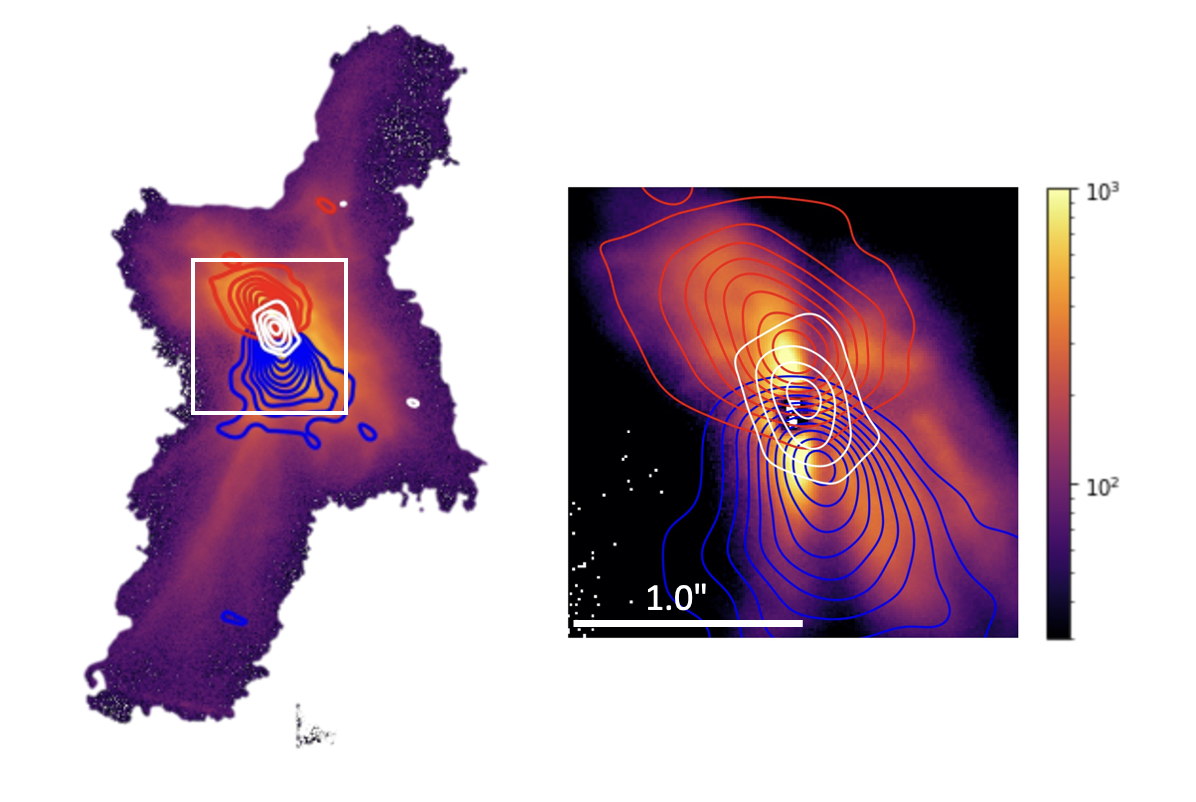} 
\caption{Overlay of SPHERE (color scale) and ALMA (contours) data. On the left all the extended structure as seen with SPHERE, on the right a zoom-in of the innermost 2$^{\prime\prime}$. White contours are ALMA 1.3 mm continuum, plotted at contours starting at, and increasing with, 3$\sigma$=0.37 mJy~beam$^{-1}$. Red and blue contours are integrated $^{12}$CO 2--1 emission over 10 km~s$^{-1}$ blue- and red-shifted relative to the source velocity, taken as V$_{\rm LSR}$=4.5 km~s$^{-1}$. Red and blue contours are also drawn starting at, and increasing with, 3$\sigma=0.12$ Jy~beam$^{-1}$~km~s$^{-1}$. The ALMA data are aligned with the SPHERE data to have the stellar position at the center of the image; the continuum emission peaks $\sim$ 0.06$^{\prime\prime}$ North of that position.}
\label{fig:ALMA_SPHERE_comparison}
\end{figure}

{\bf Feature 1.} 
The PA of the extended emission identified as feature 1 is consistent with the large scale MHOs and coincident with the small scale microjets detected through forbidden lines \citep{Pascucci2020}. Hence, we reasonably assume that it is representing outflows detected in scattered light, and that this feature is orthogonal to the inner and intermediate disk. 
The innermost disks (inner and intermediate) are misaligned with respect to the outer disk, with a PA for the inner disk of $\sim$125$^{\circ}$, measured as PA$_{\rm {outflow}}$+90$^{\circ}$. Considering the outer disk is seen with PA$_{\rm {disk}}$=7$^{\circ}$, the resulting misalignment between innermost and outer disk is of the order of 62$^{\circ}$ with an uncertainty of $\pm$10$^{\circ}$.  
This feature is illuminated by the central system. The shape of the outflow is due to higher density regions of dust, generated by instabilities created by two or more layers of material with different densities and velocities resulting in a wind-blown cavity \citep{Liang2020}. The regions with different physical properties are the highly collimated microjet (as seen from the detection of forbidden lines, e.g., \citealt{Pascucci2020}), and the surrounding wider-angle disk wind, or parent cloud. The impact between these two regions, besides carving out a large and slow massive outflow cavity into the parent cloud \citep{Frank2014}, creates regions of high density where dust grains accumulate, becoming brighter in scattered light. 
We also notice that there is a good agreement between the small scale outflow seen in the polarimetric images, and the large scale outflows determined by the MHOs, supporting the scenario of highly collimated jets carving a cavity and creating high density regions. 
We have also tested the emission seen in scattered light versus the continuum thermal emission at 1.3~mm, and the $^{12}$CO emission seen with ALMA. 
In Figure~\ref{fig:ALMA_SPHERE_comparison}, we show the continuum emission and the red- and blue-shifted line emission overlayed on the SPHERE scattered light image.
The continuum emission, shown as white contours, is slightly resolved, compact, and it is distinctly different from the orientation of the beam. 
The comparison with the SPHERE image is not quite conclusive in the direction of the emission, if along the disk or the extended emission identified as feature 1. 
$^{12}$CO line emission was clearly detected in the channels, consistent with a structure of $\sim 2.5$ arcsec in diameter. 
The emission is most likely due to the combination from emission aligned with the disk orientation inferred from SPHERE, and emission from the outflowing material in the same direction as the MHOs. 
The gas emission close to the N-S direction might trace the gas in the outer disk, and the velocity structure of the line emission is consistent with Keplerian rotation. 
The emission from the outflowing material is in the same direction as the MHOs. The velocities of the extended emission span from -3~km~s$^{-1}$ to 11~km~s$^{-1}$. The low velocities for the outflowing material confirm that the emission must happen close to the plane of the sky, as also found by \citet{Pascucci2020}. 
In both cases, either when tracing the outer disk or the outflowing material, the N-E side is receding and the S-W side is approaching the observer.

Misalignment between the inner and the outer disks are not rare. As an example, \citet{Bohn2021} have recently investigated misalignment between inner and outer disks in transitional disks, finding that out of a sample of 20 objects analyzed, six clearly show evidence of misalignment, five do not show evidence of misalignment and the others can not be evaluated with the current data. 
Misaligned disks, and disks whose orientations vary with time can be due to their formation in a turbulent, chaotic environment \citep{Bate2018}. Moreover, the evolution of the stellar and disk spin axes during the formation of a star which is accreting in a variable fashion from an inherently chaotic environment might affect the disk orientation as well \citep{Bate2010}. Also late infalling events, which carry along a specific angular momentum with respect to the star, may tilt the pre-existing disk to another rotation axis depending on the mass ratio of the mass accreted and the disk \citep{Dullemond2019, Kuffmeier2021}. This was indeed recently observed within the DESTINYS program for the SU Aur system \citep{Ginski2021}, which shows large scale streamers in scattered light, similar to those observed in our new observations of T~CrA and which were shown to trace infalling material. Stellar properties, such as strong stellar magnetic dipole, can cause a warp or misalignment in the innermost region of the disk (e.g., \citealt{Matsumoto2004, Machida2006, Matsumoto2006, Hennebelle2009, Joos2012, Krumholz2013, Li2013, Lewis2015, Lewis2017, Wurster2018}). Additionally, the presence of a companion, either stellar or substellar, can also cause inner and outer disks misalignment (e.g., \citealt{Facchini2013, Facchini2018, Zhu2019, Nealon2020}), as in the case of HD142527 \citep{Owen2017, Price2018}. 

Indeed, T~CrA and HD142527 show several similarities even if the inclinations at which the outer disks are seen are very different (almost edge-on in the case of T~CrA and almost face-on for HD142527). HD142527 is a binary system characterized by a primary 2.0~M$_{\odot}$ star surrounded by an inner disk significantly misaligned (59$^{\circ}$) with respect to the outer disk \citep{Balmer2022}. For T~CrA the outer disk is seen almost edge-on and the misalignment between outer and inner disk is coincident with the inclination of the inner disk orbit, namely $\sim$55$^{\circ}$. 
The primary star in both cases is an F-type Herbig. 
In the case of HD142527 all the main observational features (spirals, shadows seen in scattered light, horseshoe dust structure, radial flows and streamers) can be explained by the interaction between the disk and the observed binary companion \citep{Price2018}. The analysis done on HD142527 led the authors \citep{Price2018} to conclude that the disk around this Herbig star is a circumbinary rather than transitional disk, with an inclined inner disk, and with streamers of material connecting the inner and outer disk. 
In the case of T~CrA, if we assume that the inner disk is aligned perpendicular to the outflowing material, and hence misaligned with respect to the outer disk, the configuration is similar. 
Hints of dusty material inside and misaligned with respect to the outer disk come from the radial profile of the scattered light signal seen from SPHERE/IRDIS and shown in Fig.~\ref{fig:combo_rad_prof_TCrA}, where in the East-side of the disk in the direction orthogonal to the disk there is material down to the coronagraph edge. However, we cannot say from these images if this material is organized into a disk-structure itself, or if it represents a streamer of material accreting from the outer disk onto the inner regions of the system. However, as opposite to HD142527, we must mention the absence of obvious shadowing features in scattered light in T~CrA, that can nevertheless be due to the different viewing geometry. 
In the following section we will present a 3D hydrodynamical model as the one developed for HD142527 to explain the observed features as disk--binary interaction. 

{\bf Feature 2.} 
The extended emission identified as feature 2 appears very extended and resemble material falling onto the disk as in the case of SU~Aur \citep{Ginski2021}. 
Unfortunately, the strong foreground contamination due to the overall cloud does not allow to clearly detect the $^{12}$CO~(2–1), $^{13}$CO~(2–1), and C$^{18}$O~(2–1) transitions at distances larger than $\sim$2.5$^{\prime\prime}$, thus we cannot perform a detailed analysis of the kinematics of the material, as it was done, for example, in the case of SU~Aur \citep{Ginski2021}. Indeed, some parts of the CO disk may be missing from from Fig.~\ref{fig:ALMA_SPHERE_comparison} because the cloud contaminates the signal. Moreover, the large scale streamers do not show any emission due to the removing of any sensitivity to large scale emission in the data reduction process. They may exist, but they are very hard to image. The disk may also be more extended than seen here. Hence we cannot be conclusive on the nature of the extended emission in feature 2. It is highly unlikely that this emission is itself indicating outflowing material, as feature 1, but it can be most likely due to streamers of material that is falling onto the disk connecting the disk itself to the surrounding cloud material, as for SU~Aur. To some extent we might consider the scattered light morphology of T~CrA as an edge-on view of SU~Aur, where we can see the streamers of infalling material and at least one tail of accretion.  
The same streamers of accretion were already seen, but not interpreted as such, by \citet{Ward-Thompson1985, Clark2000}. \citet{Ward-Thompson1985} used linear polarization mapping of the region in R-band and identified a jet-like structure with a projected lengths of 20$^{\prime\prime}$ emerging from T~CrA, in the direction of, but pointing away from  R~CrA. \citet{Clark2000} performed near-infrared linear imaging polarimetry in J, H and Kn bands, and circular imaging polarimetry in the H band and interpreted the images as bipolar cavities, where the SE emission is visible as far as $\sim$15$^{\prime\prime}$ from T~CrA. They stress the presence of a pronounced asymmetry in the polarized intensity images, suggestive of fairly sudden depolarization of the dust grains caused by foreground material in the reflection nebula. The identification of the MHOs, and the analysis of the images acquired with NACO and SPHERE is now showing that the features observed in the past were not associated with jets but more likely the same streamer of accretion seen in scattered light.  
A possible test to ascertain the origin of feature 2 can be done using the SO$_2$ transition from ALMA. \citet{Garufi2022} have indeed shown that for the source IRAS~04302+2247, the SO$_2$ emission does not probe the disk region, but rather originates at the intersection between extended streamers and disks. 
We notice that the presence of streamers of material feeding the disk of T~CrA would also go in the direction of mitigating the issue of the low disk masses found in CrA. Indeed, it was found that the average disk mass in CrA is significantly lower than that of disks in other young (1-3 Myr) star forming regions (Lupus, Taurus, Chamaeleon I, and Ophiuchus, \citealt{Cazzoletti2019}). If there is accretion of fresh material onto the disk, one could have lower measured disk masses at the beginning, and mitigate the issue \citep{Manara2018}. The observed increase in disk masses with time (e.g., \citealt{Testi2022, Cazzoletti2019}) should otherwise be explained with other mechanisms such as planetesimal collisions \citep{Bernabo2022}.

Moreover, the presence of streamers of accretion is also in agreement with the orientation of T~CrA with respect to R~CrA, both belonging to the Coronet Cluster. These two stars formed within the same filament, which is oriented at PA=124$^{\circ}$ projected on sky (this is also the PA of T~CrA relative to R~CrA). This orientation is indeed similar to that of the orbit proposed for the central binary of T~CrA and very close to perpendicular to the PA of the MHO objects (PA=33$^{\circ}$); these values are well consistent with the direction of the same structures seen in the neighbor star R~CrA \citep{Rigliaco2019, Mesa2019}). This suggests that the bulk of the inflow of material that formed the T~CrA system was coplanar with this filament and that the original disk of T~CrA was likely oriented at the PA of the filament; this is actually the case also for the disk around R~CrA. However, the current outer disk of T~CrA has a very different orientation (PA=7 degree), though it seems to be still fed by the same filament. This is because T~CrA appears to be presently offset by a few hundreds au (a few arcsec on sky) with respect to the filament. Considering the age of T~CrA (likely 1-3 Myr), this offset is indeed very small, corresponding to a minuscule velocity of only $\sim 1$m/s. This suggests that the generation of misaligned structure is very likely whenever accretion on the disk is prolonged over such long intervals of time.

\subsection{Spectral Energy Distribution}

\begin{table}
\small
\centering
\begin{tabular}{c|c|c|c}
\hline
$\lambda_c$ & Flux & Facility & Reference  \\
($\mu$m)   & (Jy) &  &  \\\hline
       0.349  &   0.00531   & SkyMapper & \citet{Wolf2018} \\
       0.444  &   0.00792   & CTIO & \citet{Henden2016} \\
       0.444  &   0.00988   & UCAC4-RPM & \citet{Nascimbeni2016} \\
       0.482  &    0.0138   & CTIO & \citet{Henden2016} \\
       0.497  &    0.0126   & SkyMapper & \citet{Wolf2018} \\
       0.504  &    0.0142   & GAIA & \citet{GaiaEDR3} \\
       0.554  &    0.0163   & Hamilton & \citet{Herbig1988} \\
       0.554  &    0.0171   & CTIO & \citet{Henden2016} \\
       0.554  &    0.0204   & UCAC4-RPM & \citet{Nascimbeni2016} \\
       0.604  &    0.0181   & SkyMapper & \citet{Wolf2018} \\
       0.762  &     0.045   & GAIA & \citet{GaiaEDR3} \\
       0.763  &    0.0539   & CTIO & \citet{Henden2016} \\
        1.24  &     0.425   & 2MASS-J & \citet{Cutri2003} \\
        1.65  &     0.871   & 2MASS-H & \citet{Cutri2003} \\
        2.16  &      1.55   & 2MASS-K & \citet{Cutri2003} \\
        3.55  &      1.93   & Spitzer/IRAC & \citet{Gutermuth2009} \\
        4.49  &      2.07   & Spitzer/IRAC & \citet{Gutermuth2009} \\
        5.73  &      2.38   & Spitzer/IRAC & \citet{Gutermuth2009} \\
        11.6  &      3.48   & WISE/W3 &    \citet{Cutri2012} \\
        19.7  &      23.4   & SOFIA & \citet{Sandell2021} \\
        22.1  &      23.8   & WISE/W4 &    \citet{Cutri2012} \\
        25.3  &      30.7   & SOFIA & \citet{Sandell2021} \\
        31.5  &      29.0   & SOFIA & \citet{Sandell2021} \\
        37.1  &      29.3   & SOFIA & \citet{Sandell2021} \\
        70.0  &      19.3   & Herschel & \citet{HerschelCatalogue} \\
       100.0  &      14.2   & Herschel & \citet{HerschelCatalogue} \\
       160.0  &       5.0   & Herschel & \citet{HerschelCatalogue} \\
        1300  &   0.00499   & ALMA   & \citet{Cazzoletti2019} \\
        \hline
\end{tabular}
\caption{\label{tab:table_sed_tcra} List of the fluxes at different wavelengths collected from the literature used for the SED.  }
\end{table}

\begin{figure}
\center
\includegraphics[angle=0,width=0.5\textwidth]{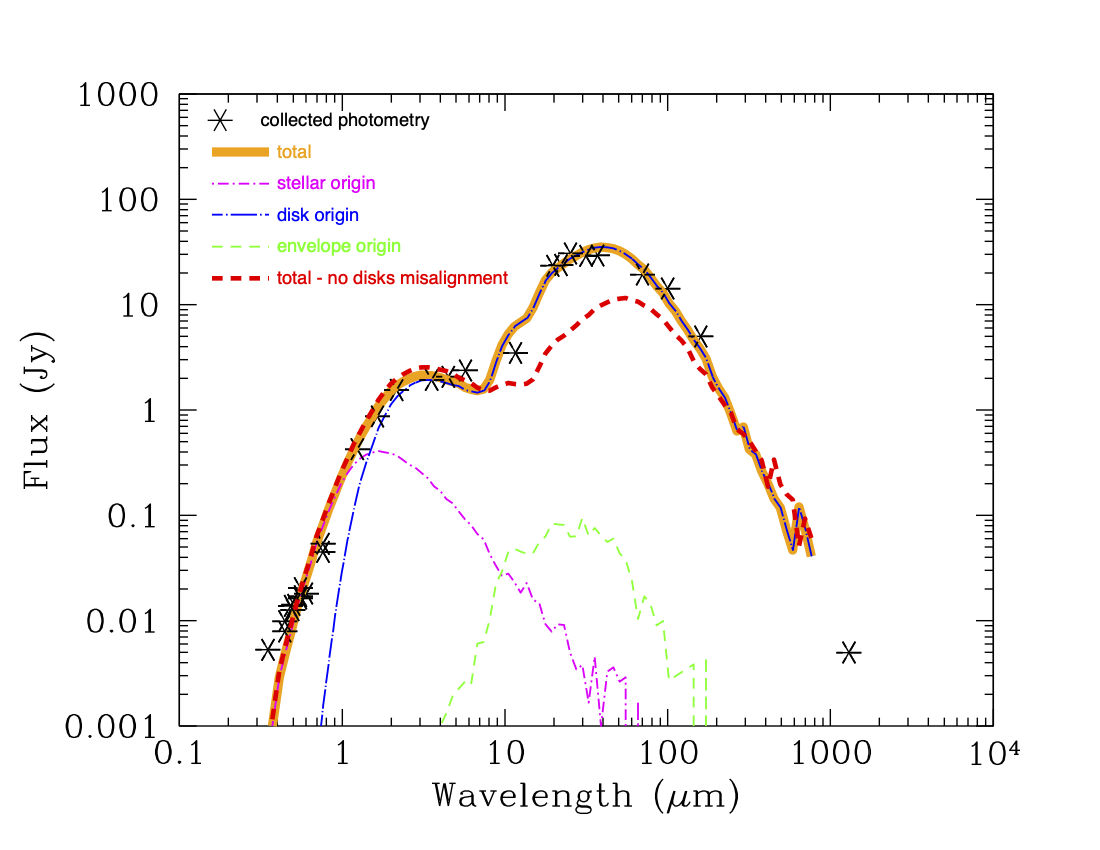} 
\caption{SED of TCrA. The black asterisks show the published photometry as reported in Table~\ref{tab:table_sed_tcra}. The orange curve shows the total emission. The magenta line shows the SED component due to stellar origin, in blue the component due to the disk, and in green the component due to the envelope. The red curve shows the emission if no misalignment between the intermediate and outer disk is assumed. The oscillations in the model curves at the longest wavelengths  are artifacts related to the finite number of photon packets considered in the Monte Carlo scheme. }
\label{fig:sed_TCrA}
\end{figure}

We model the SED of T~CrA using the dust radiative transfer model developed by \citet{Whitney2003a, Whitney2003b}.  The code uses a Monte Carlo radiative transfer scheme that follows photon packets emitted by the central star as they are scattered, absorbed, and re-emitted throughout the disk. 
For the modeling we have assumed that the geometry of the star+disk system is comprised by a central 2.0~M$_{\odot}$ source emitting photons and a gapped and misaligned circumstellar disk as described above. 
The total mass of the disk M$_{disk}$=10$^{-3}$M$_{\odot}$, which is in agreement with M$_{dust}$ retrieved by \cite{Cazzoletti2019} using the 1.3~mm continuum flux, assuming an ISM gas-to-dust ratio of 100. 
The outcome of the model, shown in orange in Fig.~\ref{fig:sed_TCrA}, well reproduces the observed photometric points collected in Table~\ref{tab:table_sed_tcra}, suggesting that the interpretation of inner and outer disks misaligned with respect to each other is in very good agreement with the collected photometry \footnote{The apparent oscillations of the model at wavelengths longer than 300~$\mu$m is due to low number statistics and has no physical meaning.}. 
For comparison, we also show the SED obtained with the same parameters, in the case where no misalignment between inner and outer disk is assumed (red profile). In this case the curve does not well reproduce the observed photometry at wavelengths longer than $\sim$10--15~$\mu$m.  
We must notice that the radiative transfer model does not account for the binary star, hence it may cause deviation in the illumination of the disk. In particular, in their orbit the two stars spend time above the disk midplane, hence illuminating the circumbinary disk from above. In the two SEDs shown in Fig.~\ref{fig:sed_TCrA} we do not account for this effect.

\subsection{Hydrodynamical Simulation} 
\label{sect:SPM_model}

We perform a 3D hydrodynamical simulation of the T~CrA configuration considered in this work using the Smoothed Particle Hydrodynamics (SPH) code {\sc Phantom} \citep{price+18, Monaghan2005, Price2012}. 
The initial conditions of the system are set following the observational constraints acquired so far. T~CrA is modeled as a binary system with masses 1.7~M$_{\odot}$, and 1.0~M$_{\odot}$ for the primary and secondary component, respectively. Each star is simulated as a sink particle \citep{price+18, bate+95} with an accretion radius of $0.5$ au. The orbit is eccentric, and the period of the binary star is 29.6~years, corresponding to a semi-major axis of 13.3~au. The orbit is seen edge-on with an inclination of 90$^{\circ}$, and PA$_{orbit}$ is perpendicular to the outflowing material (PA$_{orbit}$=145$^{\circ}$). 
The outer disk, extending from $R_{\rm in}=25$ au to $R_{\rm out}=100$ au is simulated with $8\times10^5$ SPH particles, resulting in a smoothing length $\approx0.2$ times the disk scale height. 
The inner disk, extending from $r_{\rm in}=1$~au to $r_{\rm out}=5$~au, and co-planar to the orbit of the binary star, is simulated with $2\times10^5$ SPH particles, resulting in a smoothing length of about the disk scale height. Outflows and inflows are not considered in this model. 
Viscosity is implemented with the artificial viscosity method \citep{Lucy77, Gingold&Monaghan77} that results in an \citet{Shakura&Sunyaev73} $\alpha$-viscosity as shown by \citet{Lodato&Price10}. We use $\alpha\approx5\times10^{-3}$. 
We run the full hydrodynamical model (with both the outer and the inner disk) for 100 binary orbits in order to relax the initial condition and to produce a synthetic image of the system to compare with the observation. 
To perform a direct comparison with observations of T~CrA we post-processed our simulation using the Monte Carlo radiative transfer code MCFOST \citep{Pinte2016} in order to produce synthetic images of the hydrodynamical model. MCFOST maps the physical quantities in the SPH simulation (e.g. dust and gas density, temperature) onto a Voronoi mesh directly built around the SPH particles, without interpolation.
We adopt a gas-to-dust mass ratio equals to 100 and we assume micrometer grains to be well coupled with the gas. These grains scatter the stellar light collected by SPHERE and are assumed to be spherical and homogeneous (as in the Mie theory). Their chemical composition is 60\% astronomical silicates and 15\% amorphous carbons (as DIANA standard dust composition, \citealt{Woitke2016}) and they have a porosity of 10\%. The gas mass is directly taken from the SPH simulation. We use the same distance from the source used in this paper (149.4 pc) and $\approx 10^6$ photon packets to compute the temperature profile of the model and $\approx 10^{10}$ photon packets to compute the source function of the model in order to produce the scattered light image at 2 $\mu$m wavelength. 

\begin{figure*}[h]
\center
\includegraphics[angle=0,width=1.0\textwidth]{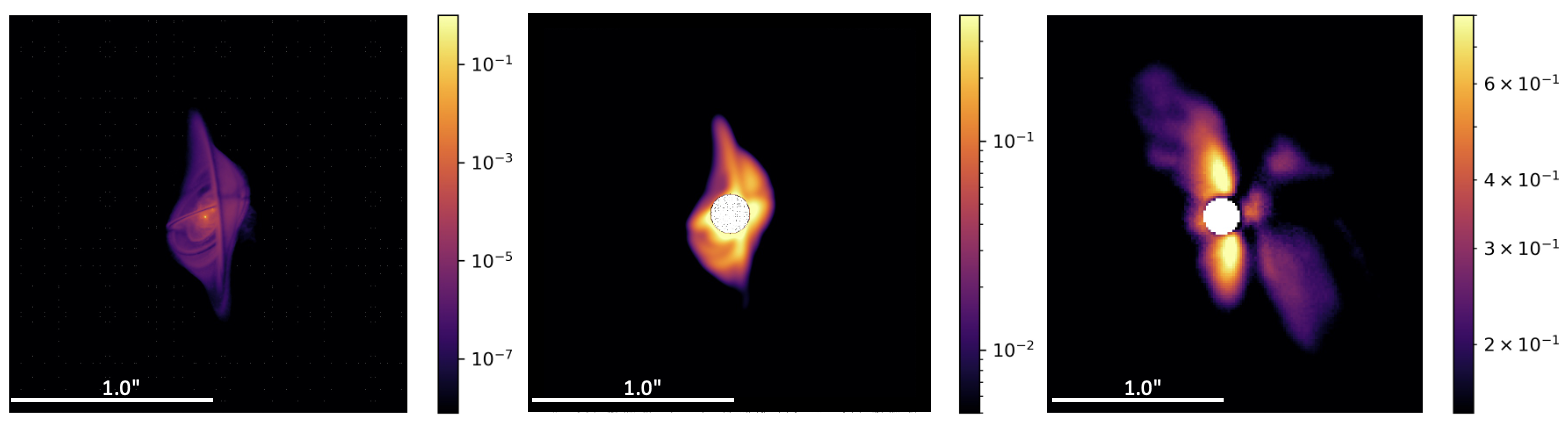}	
\caption{Snapshot of the SPH simulation compared to the observed image. The left panel shows the result in total intensity of the SPH simulation, with a resolution of 4.0 mas/pixel. In the middle panel the same image convolved to the SPHERE/IRDIS resolution (12.25 mas/pixel). On the right the observed total intensity image. All images have a 2$^{\prime\prime}$ field of view. }
\label{fig:SPH_model}
\end{figure*}

The total intensity polarized light image obtained with the hydrodynamical simulation is show in the left panel of Fig.~\ref{fig:SPH_model}. The middle panel is the synthetic image convolved to the SPHERE/IRDIS resolution and in the right panel we show the observed image. 
There are a few features that are clearly reproduced in the simulation: the dark lane, the offset of the dark lane with respect to the center of the image, the top-surface of the disk brighter than the bottom-side of the disk. 
There are two bright spots in the East-West direction on the convolved synthetic image, that are also observed in the real image. These points are due to the intermediate circumbinary disk that breaks from the outer regions, precessing as a rigid body, and leading to its evolution. 
The breaking of the inner disk generates an intermediate disk, that is visible as bright spots at the East and West side of the coronagraph.  
We must notice that the simulation does not take into consideration the outflowing material, and does not account for the replenishment of the outer disk due to the accretion streamers (hence slowing down its expansion). A more detailed simulation is needed for T~CrA, but it is beyond the scope of this observational paper and will be discussed in a separate publication. 

\begin{figure}
\center
\includegraphics[angle=0,width=0.45\textwidth]{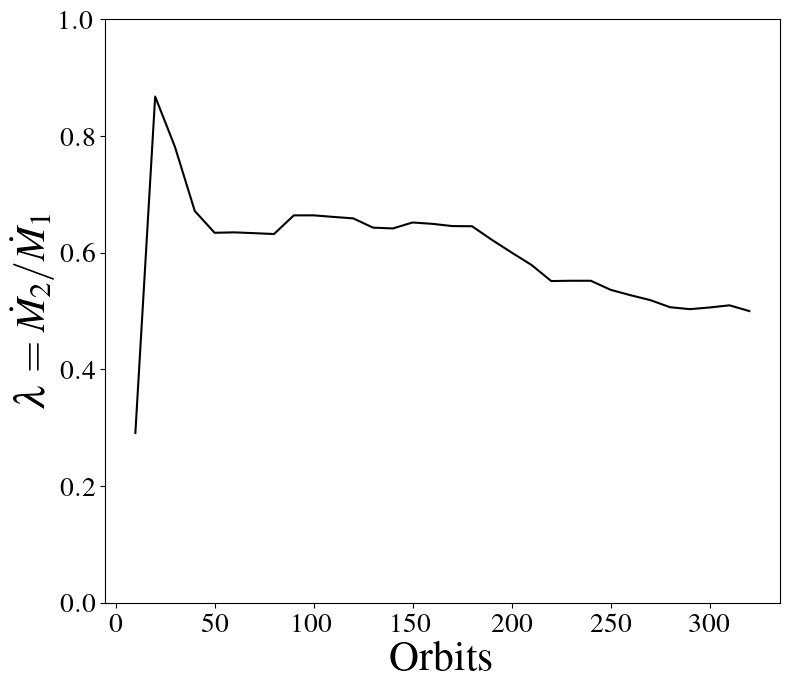}
\caption{Mass accretion rate ratio of secondary and primary star as a function of the number of orbits.  }
\label{fig:mass_ratio_SPH_model}
\end{figure}

In order to measure how the circumbinary disk mass distributes among the binary stars, we run a second hydrodynamical model as the one described above but without the circumprimary disk. Indeed, accretion into a binary system happens via the formation of up to three disks (two circumstellar disks, one around each component, and a circumbinary disk, \citet{Monin2007}). The two circumstellar disks are periodically replenished by accretion streamers pulled from the inner edge of the circumbinary disks by the stars \citep{Artymowicz1994, Tofflemire2017}. In a quasi-steady state regime, the mass flux entering the Roche lobe of a star via the gas streamers equals the star accretion rate. Thus, we can reliably measure the fraction of mass accreted onto a star by simulating only the circumbinary disk, provided that the stellar Roche lobes are resolved by the simulation and the central part of the disk has relaxed (as done with SPH simulations e.g. in \citealt{Young&Clarke15} and recently tested in \citealt{Ceppi+22}).
In general, simulations of accretion into binary systems find that the primordial mass ratio is pushed towards unity (that is, closer to equal masses in the binary components) by accretion from a circumbinary disk \citep{Clarke2012}. This is due to the ease with which the secondary component accretes the infalling gas, as it lies farther from the binary barycenter and closer to the disk edge. Its differential velocity with respect to the gas is also low, allowing it to accrete efficiently. 
In the case of T~CrA the primary star is still accreting more than the secondary (see Fig.~\ref{fig:mass_ratio_SPH_model}). This is due to the misalignment between inner and outer disk that makes the secondary to be at considerable height over/below the disk for a large fraction of its orbit.

\section{Summary and Conclusions} 

We investigate new and archival data of the Herbig Ae/Be star T~CrA collected with different instruments. The analysis of the data shows that T~CrA is a very interesting and complex system, belonging to one of the nearest and most active region of star formation. 
Combining archival NACO imaging data with photometric data, and new and archival SPHERE adaptive optics images we study the complex stellar environment around T~CrA and the stellar properties: 
\begin{itemize}
    \item{} the outer disk is seen edge-on as a dark lane elongated approximately in the N-S. The dark lane is shifted by 122~mas with respect to the center of the image, and it is seen with a PA of 7$^{\circ}$. This value is in very good agreement with the value recently found by \citealt{Cugno2022} using a different instrument and set of data; 
    \item{} the bright illuminated top-side of the disk surface is clearly visible in scattered light; 
    \item{} extended emission in the NE--SW direction, identified as feature 1, is consistent in direction with the line connecting the two-lobed MHOs seen on larger scale. It is most likely outflowing material, with PA=33$^\circ$, consistent with the PA of the two MHOs.  
    \item{} extended emission in the N-S direction, identified as feature 2, is interpreted as large scale streamers of material likely infalling onto the disk. In the North the streamer extends up to $\sim$4.5$^{\prime\prime}$ from the central system, while in the South it extends up to the edge of the field of view, and probably beyond, as suggested by previous stellar polarization images in the optical and near-IR;
    \item{} the periodic behavior of the light curve suggests a central binary with a period of 29.6~years. Even if the non-coronagraphic images acquired with NACO and SPHERE do not show direct evidence of the presence of a stellar companion, a detailed comparison of the position of the secondary along the proposed orbit at the epochs of the observations acquired so far with NACO and SPHERE shows that in all of them it was too close to the primary star for detection as a separate object. According to our modeling results the two components will be at their maximum separation in 2027: appropriate high-contrast images at that epoch should provide direct evidence of the binary system. 
\end{itemize}

Overall, we find that the binary system and intermediate circumbinary disk lay on different geometrical planes, placing T~CrA among the objects with a misaligned inner disk. 
Inner and outer disk misalignment is not rare, and in very recent years, thanks to high-contrast imaging, it is becoming clear that the misalignment can also be due to the accretion history of the star-forming cloud onto the disk. Indeed in the case of T~CrA (as well as SU~Aur) we found evidences of the presence of streamers of accreting material that connect the filament along which the star has formed with the outer part of the disk. These streamers have an angular momentum with respect to the star whose direction is very different from that of the system (in the case of T CrA, this is dominated by the binary) causing a misalignment between an inner and outer disk. 

Besides characterizing the disk/outflow structures around T~CrA, we have also modeled its spectral energy distribution, showing that the disk geometry obtained is well consistent with the observed SED, and such consistency is not reached if we do not consider the misalignment between inner and outer disk. Moreover, we have performed hydrodynamical simulation of the configuration for 100 orbits of the binary star. The model is consistent with the observations and the analysis of the accretion rates of the individual stars shows that the accretion happens mainly onto the primary star, rather than on the secondary, as a consequence of the inclination between inner/intermediate and outer disk. Also the light curve is easily explained assuming the configuration of two misaligned disks. 
Comparison of the ALMA continuum and $^{12}$CO emission have also been performed. While for the continuum emission we cannot clearly point out the region where the dust is located, if along the disk or the outflowing material, the gas emission is most likely due to the combination from emission aligned with the disk orientation inferred from SPHERE, and emission from the outflowing material in the same direction as the MHOs. 

The analysis conducted on T~CrA has confirmed its extremely interesting and complex nature. As in the case of HD142527, the misalignment between inner and outer disk can be due to the interaction between the disk and the central binary system. On the other hand, the large scale streamers observed in the N--S direction are very similar to the disk-cloud interaction observed for SU~Aur, that represents material infalling onto the disk, and inner and outer disk misalignment might be caused by this interaction. It comes clear the need for high resolution observations to disentangle the different effects that shape early planetary system formation. T CrA is an excellent target/laboratory to better understand the impact of binarity and the environment in the evolution of protoplanetary disks.

\begin{acknowledgements}
We would like to thank the referee Roubing Dong, whose careful and constructive comments improved the quality of this manuscript.
E.R. was supported by the European Union’s Horizon 2020 research and innovation programme under the Marie Skłodowska-Curie grant agreement No 664931. This work has been supported by the project PRIN INAF 2016 The Cradle of Life - GENESIS-SKA (General Conditions in Early Planetary Systems for the rise of life with SKA) and by the "Progetti Premiali" funding scheme of the Italian Ministry of Education, University, and Research. 
C.F.M acknowledges funding from the European Union under the European Union’s Horizon Europe Research \& Innovation Programme 101039452 (WANDA). Views and opinions expressed are however those of the author(s) only and do not necessarily reflect those of the European Union or the European Research Council. Neither the European Union nor the granting authority can be held responsible for them. 
T.B. acknowledges funding from the European Research Council (ERC) under the European Union's Horizon 2020 research and innovation programme under grant agreement No 714769 and funding by the Deutsche Forschungsgemeinschaft (DFG, German Research Foundation) under grants 361140270, 325594231, and Germany's Excellence Strategy - EXC-2094 - 390783311. 
A.R. has been supported by the UK Science and Technology research Council (STFC) via the consolidated grant ST/S000623/1 and by the European Union’s Horizon 2020 research and innovation programme under the Marie Sklodowska-Curie grant agreement No. 823823 (RISE DUSTBUSTERS project). 
This paper makes use of the following ALMA data: ADS/JAO.ALMA\#2016.0.01058.S. ALMA is a partnership of ESO (representing its member states), NSF (USA) and NINS (Japan), together with NRC (Canada), MOST and ASIAA (Taiwan), and KASI (Republic of Korea), in cooperation with the Republic of Chile. The Joint ALMA Observatory is operated by ESO, AUI/NRAO and NAOJ. MRH acknowledges the assistance of Allegro, the ARC node in the Netherlands, who assisted with the calibration of this data set.
This work is partly based on data products produced at the SPHERE Data Centre hosted at OSUG/IPAG, Grenoble. We thank P. Delorme and E. Lagadec (SPHERE Data Centre) for their efficient help during the data reduction process. SPHERE is an instrument designed and built by a consortium consisting of IPAG (Grenoble, France), MPIA (Heidelberg, Germany), LAM (Marseille, France), LESIA (Paris, France), Laboratoire Lagrange (Nice, France), INAF Osservatorio Astronomico di Padova (Italy), Observatoire de Genève (Switzerland), ETH Zurich (Switzerland), NOVA (Netherlands), ONERA (France) and ASTRON (Netherlands) in collaboration with ESO. SPHERE was funded by ESO, with additional contributions from CNRS (France), MPIA (Germany), INAF (Italy), FINES (Switzerland) and NOVA (Netherlands). SPHERE also received funding from the European Commission Sixth and Seventh Framework Programmes as part of the Optical Infrared Coordination Network for Astronomy (OPTICON) under grant number RII3-Ct-2004-001566 for FP6 (2004-2008), grant number 226604 for FP7 (2009-2012), and grant number 312430 for FP7 (2013-2016). 
This work has made use of data from the European Space Agency (ESA) mission
{\it Gaia} (\url{https://www.cosmos.esa.int/gaia}), processed by the {\it Gaia}
Data Processing and Analysis Consortium (DPAC,
\url{https://www.cosmos.esa.int/web/gaia/dpac/consortium}). 
Funding for the DPAC has been provided by national institutions, in particular the institutions participating in the {\it Gaia} Multilateral Agreement.
We acknowledge with thanks the variable star observations from the AAVSO International Database contributed by observers worldwide and used in this research.
\end{acknowledgements}

\bibliographystyle{aa} 
\bibliography{sample.bib} %

\begin{appendix}

\section{SPHERE polarimetric images}
\label{sect:AppA}

\begin{figure}[!h]
\center
\includegraphics[angle=0,width=0.5\textwidth]{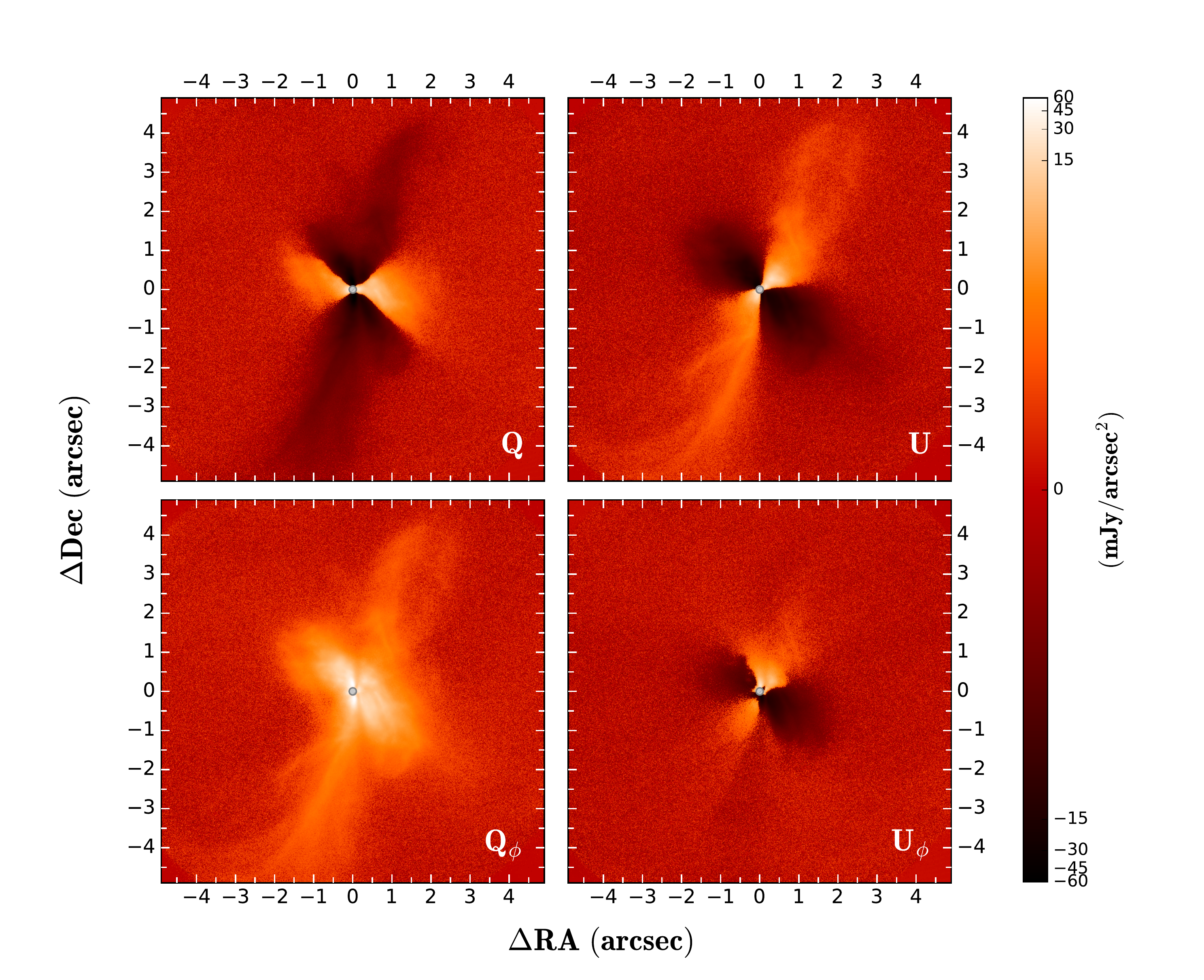}
\caption{Flux calibrated image of the Q, U, Q$_{\Phi}$ and U$_{\Phi}$ frames. }
\label{fig:pol_fl_cal}
\end{figure}

In Figure~\ref{fig:pol_fl_cal} we present the Stokes Q and U, as well as the derived Q$_{\Phi}$ and U$_{\Phi}$ images of T~CrA. The flux calibration was carried out by measuring the flux of the central star in the non-coronagraphic flux calibration images, taken at the beginning and end of the observation sequence. To convert pixel counts to physical units we used the 2MASS H-band magnitude of the star. 

\section{Proper motion analysis}
\label{sect:AppB}

The average proper motion for the on-cloud Coronet cluster members obtained using Gaia DR2 data from \citealt{Galli20}, is $\mu_{\alpha}\cos{\delta}=4.3$~mas~yr$^{-1}$ and $\mu_{\delta}$=-27.3~mas~yr$^{-1}$ with a small dispersion of less than 1 mas/yr for the individual objects. As we mentioned in the introduction, Gaia does not provide astrometric solutions and proper motion for the star T~CrA. However, we can check for peculiar/transient motion of the star using the following procedure. We collect from UCAC4 \citep{Zacharias2012}, PPMXL \citep{Roeser2010} and Gaia DR3 \citep{Gaia2016, Gaia2020} database a list of 10 bright stars in the T~CrA surroundings for which proper motion are available in all catalogs. These stars, listed in Table~\ref{tab:PM_analysis}, are selected such that they have mag$_G\leq$14 and lay within 10$^{\prime}$ from T~CrA. For these stars we measure a long term proper motion given by the difference in position between the UCAC4, PPMXL, and Gaia DR3 epochs. 
The long term proper motion, defined as the motion of the star between different epochs of observations is measured as: 
\begin{equation}
    \mu_{\alpha}\cos{\delta}=\frac{(RA_{Epoch 1}-RA_{Epoch 2})*cos(DEC_{Epoch 1})}{(Epoch 1-Epoch 2)}  
\end{equation}
\begin{equation}
    \mu_{\delta}=\frac{(DEC_{Epoch 1} - DEC_{Epoch 2})}{(Epoch1 - Epoch2)}
\end{equation} 
The analysis of the proper motion between the various epochs of the selected stars allows us to find a systematic offset between the average of the coordinate systems of UCAC4 and PPMXL with respect to Gaia DR3, that averages to 0.41$\pm$2.46~mas~yr$^{-1}$ in RA and 9.67~mas~yr$^{-1}\pm$2.93~mas~yr$^{-1}$ in DEC for this specific region of the sky. 
For T~CrA we obtain an estimate of the proper motion of the star by correcting the long term proper motion with the systematic offset, finding as values $\mu_{\alpha}\cos{\delta}=8.5\pm 2.5$~mas~yr$^{-1}$ and $\mu_{\delta}$=-33.5$\pm$2.9~mas~yr$^{-1}$. 
Considering the average proper motion of the on-cloud members we find for T~CrA an apparent motion $\mu_{\alpha}\cos{\delta}=4.2\pm 2.5$~mas~yr$^{-1}$ in RA and $\mu_{\delta}$=-6.2$\pm$2.9~mas~yr$^{-1}$ in DEC.

\begin{figure}[h]
\center
\includegraphics[angle=0,width=0.45\textwidth]{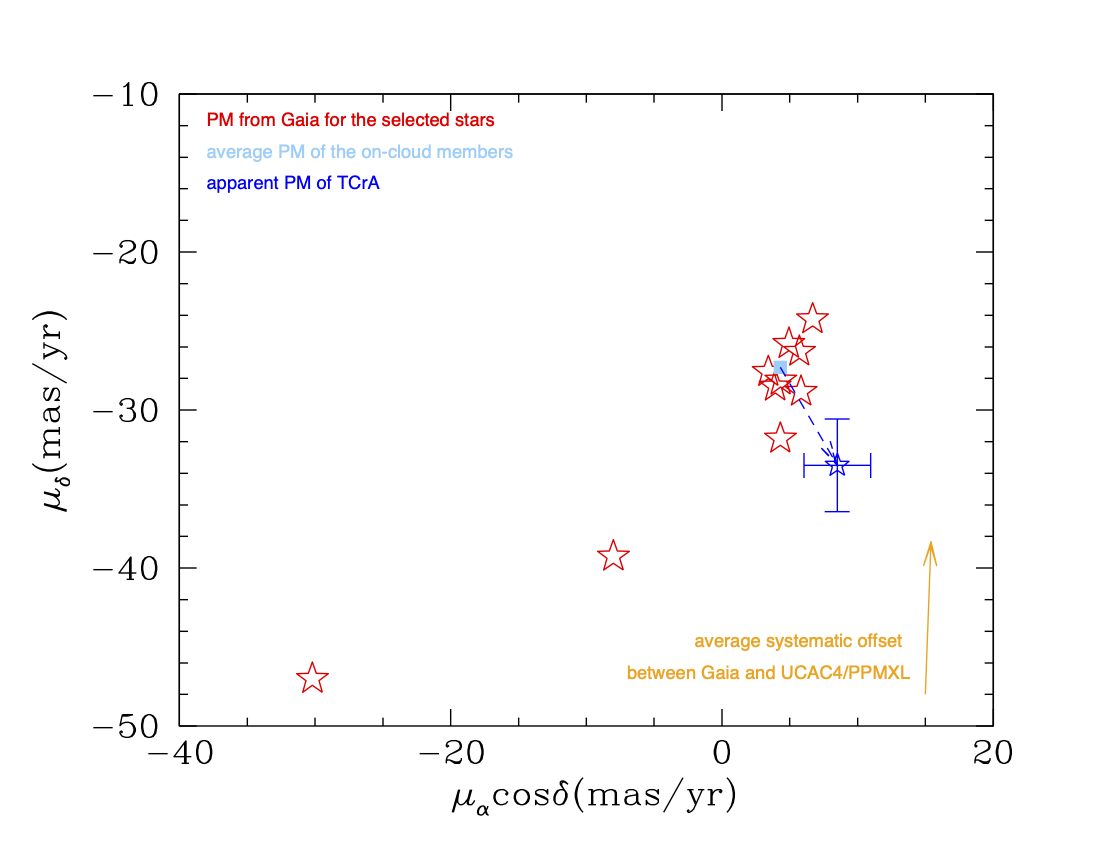}
\caption{Proper motion of the ten stars reported in Table~\ref{tab:PM_analysis}. The cyan square represents the average proper motion for the on-cloud Coronet cluster obtained using Gaia DR2 data \citep{Galli20}. The blue star is the calculated apparent proper motion of T~CrA after correcting the long term proper motion for the systematic offset. In orange the direction of the systematic offset of the proper motion due to the different coordinate system.  }
\label{fig:pm}
\end{figure}

\begin{landscape}
\begin{table}
\small
\begin{tabular}{l|c|c|c|c|c|c|c|c|c|c|c|c|c}
\hline
& T~CrA & V709 CrA & HD176269 & HD176270 & TY CrA & HD176423 & V702 CrA & HD176386 & HD176497 & HD176018 & CD-36 13202\\
\hline 
\multicolumn{12}{c}{(UCAC4)} \\
RA & 285.494904 & 285.395229 & 285.263532 & 285.267908 & 285.420107 & 285.460076 & 285.508229 & 285.412217	& 285.528297 & 284.930902 & 284.920385			 \\ 
Dec &	-36.963871 & -37.015723 & -37.060898 & -37.061555 & -36.876063 & -36.664651 & -37.128761 & -36.890712 & -36.361622 & -36.788004 & -36.588797 \\
Ep. RA & 1997.40 & 1985.45 & 1991.25 & 1991.25 & 1991.09 & 1990.50 & 1985.88 & 1991.25 & 1990.57 & 1988.91 & 1995.62 \\
Ep. Dec & 1997.77& 1985.43 & 1991.25 & 1991.25 & 1990.52 & 1989.91 & 1984.44 & 1991.25 & 1990.28 &1988.04 & 1995.74 \\
\hline
\multicolumn{12}{c}{(PPMXL)} \\
RA & 285.494908 & 285.395224 & 285.263532 & 285.267908 & 285.420102	& 285.460076 & 285.508229 & 285.412219 & 285.528303 & 284.930902 & 284.920394 \\ 
Dec & -36.963869 & -37.015722 & -37.060898 & -37.061555 & -36.876064 & -36.664651 & -37.128761 & -36.890703 & -36.361625 & -36.788007 & -36.588800 \\
Ep. RA & 1999.95 & 1988.00 & 1991.73 & 1991.18 & 1991.53 & 1991.41 & 1997.44 & 1991.14 & 1991.32 & 1991.23 & 1997.69 \\
Ep. Dec & 1999.95 & 1986.70 & 1991.64 & 1991.19 & 1991.76 & 1991.62 & 1998.14 & 1991.09 & 1991.38 & 1991.64 & 1998.44 \\
\hline
\multicolumn{12}{c}{(Gaia DR3)} \\
RA & 285.494959 & 285.395278 & 285.263578 & 285.267966 & 285.420142 & 285.460103 & 285.508268 & 285.412238 & 285.528324 & 284.930856 & 284.920226 \\ 
Dec & -36.963983 & -37.015845 & -37.061040 & -37.061682 & -36.876201 & -36.664770 & -37.128870 & -36.890838 & -36.361752 & -36.788188 & -36.589008\\
Ep. RA & 2016.0 & 2016.0 & 2016.0 & 2016.0 & 2016.0& 2016.0 & 2016.0 & 2016.0 & 2016.0 & 2016.0 & 2016.0 \\
Ep. Dec & 2016.0 & 2016.0 & 2016.0 & 2016.0 & 2016.0 & 2016.0 & 2016.0 & 2016.0 & 2016.0 & 2016.0 & 2016.0 \\
\hline
\hline
\end{tabular}
\caption{\label{tab:PM_analysis} List of the stars used to measure the proper motion offset.}
\end{table}
\end{landscape}

\section{Binarity and light curve}
\label{sect:AppD}

The light curve of T~CrA appears to be periodic. The period is found to be 29.6~years and it can be due to the presence of a binary star at the center of the T~CrA system with a mass ratio q$\sim$0.5$\pm$0.2, that is partially obscured by a disk seen edge-on, that has an offset with respect to the photocenter of the binary star of $\sim$90~mas. The model of this binary system, described in Sect~\ref{sect:binarity}, is also able to account for the large apparent proper motion measured in the period between 1998 and 2016. None of the images acquired in recent years with NACO (in 2007, 2016 and 2017) and SPHERE (in 2016, 2018 and 2021) shows clear evidence of a binary system for T~CrA. Hence, we have checked what was the relative position of the secondary star with respect to the primary for every single epoch for which we have an image, and the H-band contrast that should be observed. These quantities are shown in Table~\ref{tab:rel_motion_contrast}. 
These value are all consistent with the fact that the binary system is not clearly resolved. Indeed, in the 2007, 2018 and 2021 epochs the separation between the two stars is too small to see the two sources separately. On the contrary, the two 2016 epochs have a larger separation, though still within 2$\times \lambda/D$, that is so close that the secondary cannot be clearly separated from the primary. We notice however, that in both images acquired around this epoch with NACO and SPHERE, the PSF appears elongated in the NW-SE direction, that corresponds to the direction of the major axis of the orbital motion of the binary system. The average position angle of the elongated PSFs acquired in 2016 is 130$\pm$15$^{\circ}$, in very good agreement with the direction of the peculiar proper motion (PA$_{PM}$) measured, and with the hypothesis that the orbit of the binary system is seen edge-on, and perpendicular to the outflow. This elongation in different epochs supports then the scenario of a binary star. 
In a few years, namely in 2027, when the system is at its highest separation, the secondary component should be detectable with high-contrast images. 

\begin{table}
\small
\centering
\begin{tabular}{c|c|c}
\hline
Epoch &  Offset B-A (mas) & dH (mag) \\
\hline
2007.54 (NACO) &   -26.0$\pm$7.0   &  1.0$\pm$0.6 \\
2016.25 (NACO) &   -72.0$\pm$5.0   &  0.0$\pm$0.7 \\ 
2016.60 (SPHERE) & -69.0$\pm$5.0 & 0.2$\pm$0.7 \\
2018.36 (SPHERE )& -44.0$\pm$7.0 & 0.3$\pm$0.6 \\
2021.50 (SPHERE) &    11.0$\pm$7.0   &  1.0$\pm$0.5 \\
\hline
\end{tabular}
\caption{\label{tab:rel_motion_contrast} Relative position of the secondary star (B) with respect to the primary star (A), and relative contrast (dH) in H-band of the secondary star with respect to the primary star, for a period of 29.6~years. The offset is defined in the direction of the semi-major axis of the stellar orbit. 
}
\end{table}

\begin{figure}[h]
\center
\includegraphics[angle=0,width=0.45\textwidth]{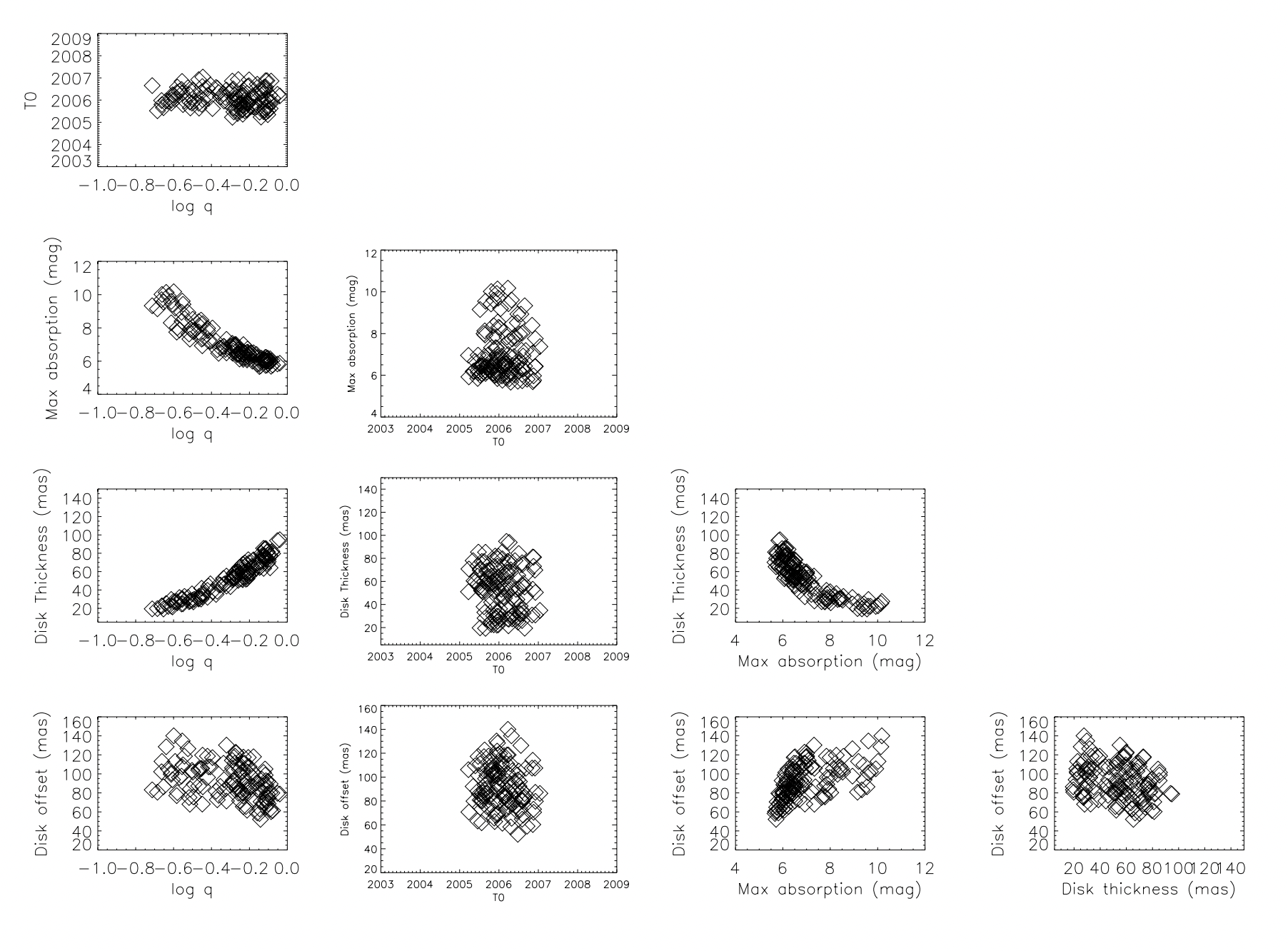}
\caption{ Corner plot showing the results of the MC parameters estimation for the model described in the paper when a period of 29.2~years is considered. The plots show the 2D joints posterior densities of all couple of parameters. }
\label{fig:cp29}
\end{figure}

\begin{figure}[h]
\center
\includegraphics[angle=0,width=0.45\textwidth]{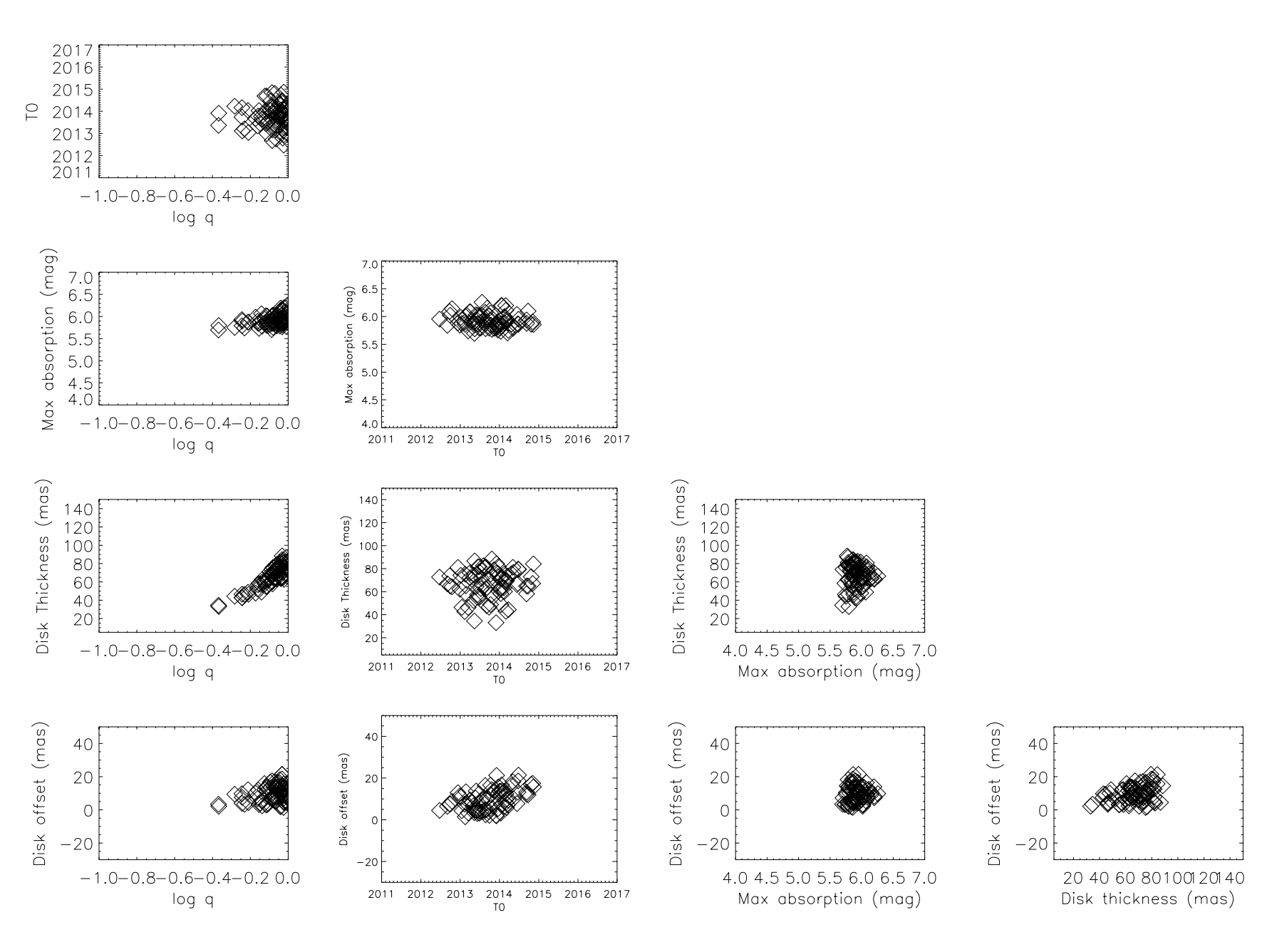}
\caption{ Corner plot showing the results of the MC parameters estimation for the model described in the paper when a period of 59.6~years is considered. The plots show the 2D joints posterior densities of all couple of parameters. }
\label{fig:cp59}
\end{figure}

\begin{figure}[h]
\center
\includegraphics[angle=0,width=0.45\textwidth]{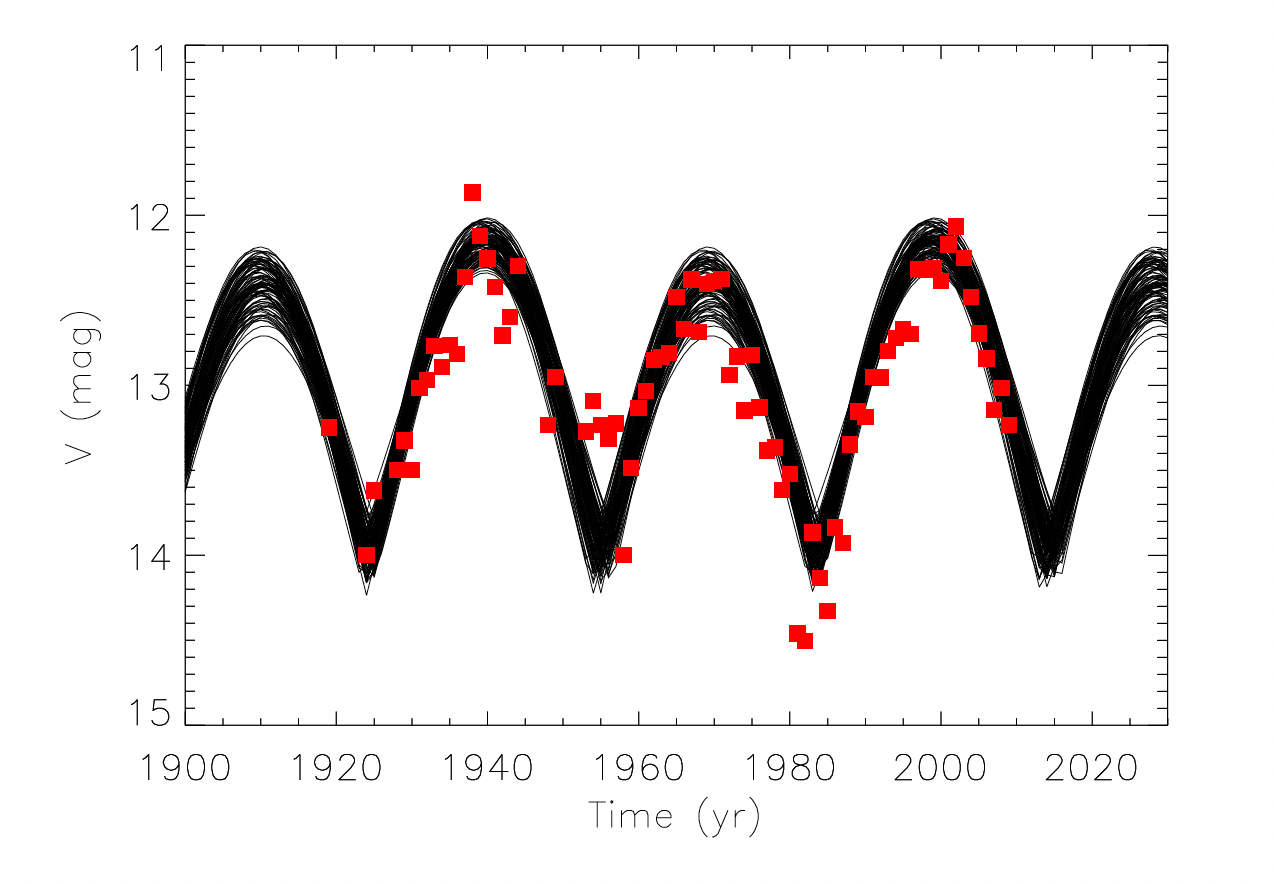}
\caption{Light curve of T~CrA (red points) compared to the light curves computed with the MC model (black lines) assuming a period of 59.2~years.   }
\label{fig:lc59}
\end{figure}

We have also considered that the period of the system might be double than the period measured in Sect.~\ref{sect:photom_data}, namely 59.2~years. While the light curve can be, also in this case, easily reproduced, there are several observational shortcomings in this interpretation. First of all we must notice that in this case the model predicts a mass ratio q as high as 0.9, and an offset of the disk of $\sim$10~mas. This last quantity is in disagreement with the observations, that instead show that the disk dark lane has an offset ten times larger. Moreover, the position of the center of the binary system as retrieved by assuming a 59.2~years period is not consistent with the motion of the system obtained from UCAC4/PPMXL and Gaia DR3 data. Additionally, in the SPHERE image acquired in 2021, the predicted separation between the primary and secondary component should be 108$\pm$6~mas, with a contrast dH=0.5$\pm$0.1~mag, making it visible as a separate point source in the image. The relative position of the secondary star with respect to the primary for every single epoch for which we have an image, and the H-band contrast that should be observed are reported in Table~\ref{tab:rel_motion_contrast_59}. 
The image does not reveal the presence of the secondary star. Given these shortcomings between observations and the output of the model, we exclude that the period of the binary star is 59.2~years. 
Figure~\ref{fig:cp29} and ~\ref{fig:cp59} show the corner plot of the derived quantities of the model used in Sect.~\ref{sect:mod_lc} to model the light curve assuming a period of 29.6~years or the double (59.2~years). 
The light curve for the period of 59.2~years is shown in Fig.~\ref{fig:lc59}. 

\begin{table}
\small
\centering
\begin{tabular}{c|c|c}
\hline
Epoch &  Offset B-A (mas) & dH (mag) \\
\hline
2007.54 (NACO) &      91.0$\pm$7.0 & 1.0$\pm$0.2 \\
2016.25 (NACO) &    -39.0$\pm$8.0 & 0.0$\pm$0.3 \\
2016.60 (SPHERE) &     -44.0$\pm$8.0 & 0.0$\pm$0.2 \\
2018.36 (SPHERE) &     -70.0$\pm$7.0 & 0.0$\pm$0.2 \\
2021.50 (SPHERE) &     -108.0$\pm$6..0 & 0.0 $\pm$ 0.1 \\
\hline
\end{tabular}
\caption{\label{tab:rel_motion_contrast_59} Relative position of the secondary star (B) with respect to the primary star (A), and relative contrast (dH) in H-band of the secondary star with respect to the primary star, for a period of 59.2~years. The offset is defined in the direction of the semi-major axis of the stellar orbit. 
}
\end{table}

\end{appendix}

\end{document}